\documentclass[a4paper,onecolumn,
superscriptaddress,11pt,accepted=2019-06-24]{quantumarticle}
\pdfoutput=1
\usepackage[utf8]{inputenc}
\usepackage[english]{babel}

\usepackage{tikz}
\usepackage{lipsum}

\usepackage{amsmath, amssymb, amsfonts, amsthm}
\usepackage{%natbib, 
hyperref, doi}
\usepackage{xcolor} 
\newcommand{\blue}{}

\newcounter{thaler}
\newenvironment{mlist}{\begin{list}{\arabic{thaler}}%
{\usecounter{thaler}
\setlength{\rightmargin}{\leftmargin}
\topsep=0pt
\itemsep=0pt
\parskip=0pt
\parsep=0pt
}}{\end{list}}

\newcommand{\tempout}[1]{{}}

\usepackage{tikz}
\tikzset{help lines/.style=very thin}

\usepackage[arrow,matrix,curve,cmtip,ps]{xy}

\theoremstyle{plain}
\newtheorem{theorem}{Theorem}%[section]
\newtheorem{lemma}{Lemma}

\newtheorem{corollary}{Corollary}

\theoremstyle{definition}

\newtheorem{definition}{Definition}

\newcommand{\jProd}{\raisebox{-0.88ex}{\scalebox{2.5}{$\cdot$}}}
\renewcommand{\dot}{\jProd}

\newcommand{\Proc}{\mbox{Proc}}
\newcommand{\Aut}{\mbox{Aut}}

\newcommand{\Tr}{\mbox{Tr}}

\newcommand{\M}{{\mathcal M}}

\newcommand{\E}{{\mathbf E}}
\newcommand{\V}{{\mathbf V}}
\newcommand{\J}{{\boldsymbol E}}
\renewcommand{\Vec}{{\boldsymbol E}}
\newcommand{\F}{{\boldsymbol F}}
\renewcommand{\J}{{\boldsymbol J}}
\newcommand{\B}{{\mathcal B}}

\newcommand{\R}{{\mathbb R}}
\newcommand{\C}{{\mathcal C}}

\newcommand{\x}{{\mathbf x}}
\newcommand{\y}{{\mathbf y}}
\renewcommand{\H}{\boldsymbol{\mathcal H}}
\newcommand{\K}{\boldsymbol{\mathcal K}}
\newcommand{\id}{\mbox{id}}
\renewcommand{\1}{\boldsymbol{1}}
\newcommand{\0}{\boldsymbol{0}}
\renewcommand{\L}{{\mathcal L}}
\renewcommand{\bar}{\overline}
\renewcommand{\hat}{\widehat}

\newcommand{\sa}{\mbox{\tiny sa}}

\renewcommand{\tilde}{\widetilde}
\renewcommand{\hat}{\widehat}
\newcommand{\spn}{\mbox{Span}}

\newcommand{\Cat}{{\mathcal C}}

\newcommand{\red}{}%\color{red}}
\tempout{
 \title{\bf Conjugates, Filters and Quantum Mechanics}
\author{Alexander
  Wilce \footnote
  {Department of Mathematics and Computer Science, 
  Susquehanna University {\tt wilce@susqu.edu}}}
\date{June 6, 2018} 
}

\begin{document}

\title{Conjugates, Filters and Quantum Mechanics}
\date{\today}
\author{Alexander Wilce}
\affiliation{Department of Mathematics and Computer Science, Susquehanna University}
\email{wilce@susqu.edu}
%\homepage{}
%\orcid{0000-0003-0290-4698}
%\thanks{I wish to thank Giulio Chiribella, Chris Heunen, Matt Leifer and Markus M\"{u}ller for helpful comments on earlier %drafts
%of this paper. This work was  supported in part by a grant (FQXi-RFP3-1348) from the FQXi foundation.}. 

\maketitle

\begin{abstract} 
The Jordan structure of finite-dimensional quantum theory is derived, in a conspicuously easy way, from a few simple postulates concerning abstract probabilistic models (each defined by a set of basic measurements and a convex set of states).  The key assumption is that each system $A$ can be paired with an isomorphic \emph{conjugate} system, 
$\overline{A}$, by means of a non-signaling bipartite state $\eta_A$ perfectly and uniformly correlating each basic measurement on \(A\) with its counterpart on $\overline{A}$. In the case of a quantum-mechanical system associated with a complex Hilbert space \(\H\), the conjugate system is that associated with the conjugate Hilbert space $\H$, and $\eta_A$ corresponds to the standard maximally entangled EPR state on $\H \otimes \overline{\H}$.  A second ingredient is the notion of a {\em reversible filter}, that is, a probabilistically reversible process that independently attenuates the sensitivity of detectors associated with a measurement. In addition to offering more flexibility than most existing reconstructions of finite-dimensional quantum theory, the approach taken here has the advantage of not relying on any form of the ``no restriction" hypothesis. That 
is, it is not assumed that arbitrary effects are physically measurable, nor that arbitrary families of physically measurable effects summing to the unit effect, represent physically accessible observables. (An appendix shows  
how a version of Hardy's ``subpace axiom" can replace several assumptions native to this paper, although at the 
cost of disallowing superselection rules.)
\end{abstract}

\section{Introduction and Overview} 

A number of recent papers, notably \cite{CDP, Dakic-Brukner, Hardy, Masanes-Mueller, Rau}, have succeeded in deriving the mathematical apparatus of finite-dimensional quantum mechanics (henceforth: QM) from various packages of broadly operational, probabilistic, or information-theoretic assumptions. 
These assumptions are, however, rather strong, and the derivations themselves are not trivial. This paper aims at a slightly broader target, and finds it much easier to hit. 

Specifically, the Jordan structure of finite-dimensional quantum theory is derived, in a conspicuously easy way, from a few simple principles. This still brings us within hailing distance of standard QM, owing to the classification theorem 
for finite-dimensional formally real Jordan algebras as direct sums of real, complex and quaternionic quantum systems, spin factors (``bits" of arbitrary dimension), and the exceptional Jordan algebra \cite{JNW}. 
In contrast, all of the cited reconstructions make use of strong axioms that rule out real and quaternionic 
systems, and even complex quantum systems with superselection rules, more or less by {\em fiat}. Since there 
are good arguments for taking real and quaternionic quantum systems seriously (see \cite{Baez} for a 
forceful argument in this direction), it is of interest to have an axiomatic scheme that accommodates them. 
I shall have more to say on this point below.\\

\noindent{\bf Correlation in quantum mechanics} The approach taken here  begins with a simple and well-known observation about finite-dimensional quantum systems. 
Let $\H$ be an $n$-dimensional complex Hilbert space\footnote{A word on notation: I follow the mathematicians' convention that a complex inner product $\langle \ , \ \rangle$ is conjugate-linear in the second, rather than the first argument. 
Thus, in terms of Dirac notation, $\langle x , y \rangle = \langle y | x \rangle$.}, representing a 
finite-dimenional quantum system. 
{\blue Recall that the {\em conjugate Hilbert space}, $\bar{\H}$, % be the conjugate Hilbert space. 
%That is, $\bar{\H}$ 
is the same abelian group, but endowed with the scalar multiplication 
$(c,x) \mapsto \bar{c}x$ (where the scalar multiplication on the right is that in $\H$, and 
$\bar{c}$ is the complex conjugate of $c \in \C$), and with inner product 
$(x,y) \mapsto \langle y, x \rangle$. It is customary 
to write $\bar{x}$ for the vector $x \in \H$, {\em regarded as} a vector in $\bar{\H}$, so that 
$c \bar{x} = \bar{c}x$, or, equivalently, $\bar{cx} = \bar{c} \, \bar{x}$. The inner product on 
$\H$ is then given by $\langle \bar{x}, \bar{y} \rangle = \langle y, x \rangle = \bar{\langle x, y \rangle}$.
\footnote{One can think of $\bar{\H}$ as the space of bras $\langle x |$ corresponding to the kets $|x\rangle \in \H$, but 
I prefer to avoid this representation, since I want to stress the idea that $\bar{\H}$ represents 
a quantum system in its own right. Thus, using Dirac notation we might write 
$|\bar{x} \rangle = \langle x |$.  }}

Suppose now that $W$ is any density operator on $\H$, with spectral decomposition 
$W = \sum_{x \in E} \lambda_x p_x$ for some orthonormal basis $E$, where $p_x$ is the rank-one projection associated with a unit vector $x \in E$. Then $W$ is the marginal, or reduced state, 
of the pure bipartite state 
\begin{equation}
\Psi_{W} \ := \ \sum_{x \in E} \lambda^{1/2}_{x} x \otimes \bar{x} \ \in  \H \otimes \bar{\H}
\end{equation}
The fact that mixed quantum-mechanical states 
arise in this way as marginals of pure states on larger systems is the starting point for the reconstruction of QM 
in \cite{CDP}. Here, we focus instead on the correlational features of $\Psi_W$. 
A straightforward calculation shows 
that if $a,b$ are any two operators on $\H$, then  
\[\langle (a \otimes \bar{b}) \Psi_{W}, \Psi_{W} \rangle  = \Tr(W^{1/2}aW^{1/2}b).\]
In particular, if $a$ and $b$ commute with $W$, then we have 
\begin{equation}
%\[
\langle (a \otimes \bar{b}) \Psi_{W} , \Psi_{W} \rangle = \Tr(Wab).%\] 
\end{equation}
It follows that the state $\Psi_W$ 
{\em perfectly correlates} any projection-valued observable that commutes with $W$, with its counterpart on $\bar{\H}$: if $a$ and $b$ are mutually orthogonal 
projections, both commuting with $W$, then the joint probability of observing $a$ and $\bar{b}$ is $\langle \Psi_{W}, a \otimes \bar{b} \rangle  = \Tr(Wab) = 0$, 
while the joint probability of $a$ and $\bar{a}$  is $\langle \Psi_{W}, a \otimes \bar{a} \rangle = \Tr(Wa)$.
Where  %$a = |x\rangle \langle x |$ %
$a = p_{x}$ 
is the rank-one projection associated with a unit vector $x$, this means that the conditional state of the conjugate system, given a measurement result $x$ on the system corresponding to $\H$, is the 
``collapsed" state corresponding $\bar{x}$. In effect, the entangled state $\Psi_W$ allows 
the conjugate system to retain a {\em record} of the measurement result on the first system --- even though no signal need have passed between the two.

A striking special case arises where $W = \tfrac{1}{n} \1$, the {\em \blue maximally mixed} state: in this case, 
$\Psi_{W}$ is the ``EPR" state 
\[\Psi \ = \ \tfrac{1}{\sqrt{n}} \sum_{x \in E} x \otimes \bar{x},\]
the expansion being independent of the choice of the orthonormal basis $E$. As {\em every} observable commutes with $W$, $\Psi$ perfectly, and {\em uniformly}, 
correlates {\em every}  observable on $\H$ with its counterpart on $\bar{\H}$. Thus, if we imagine that 
the system corresponding to $\H$ is controlled by Alice and that corresponding to $\bar{\H}$, by Bob,
then if Alice and Bob happen to make the same measurement, they are bound to obtain the same result, 
with uniform probability $1/n$. 
Notice, also, that  by (2) we have 
\[\langle (a \otimes \bar{b}) \Psi, \Psi \rangle = \tfrac{1}{n} \Tr(ab)\] 
for all observables $a$ and $b$, 
so the state $\Psi$ in some sense {\em explains} the normalized trace inner product. \\

\noindent{\bf Correlation in General Probabilistic Theories} These correlational features make sense in a much more 
general setting. As explained in more detail below, a {\em probabilistic model}  is characterized by a set of basic measurements or experiments, and a convex set of states, with each state $\alpha$ assigning a probability $\alpha(x)$ to every outcome $x$ of every basic measurement. 
Given two such models $A$ and $B$, a {\em bipartite state} $\omega$ on $A$ and $B$ is an assignment of joint probabilities $\omega(x,y)$ to all outcomes $x$ and $y$ of basic $A$- and $B$-measurements, respectively, having well defined conditional and marginal (reduced) probability weights corresponding to states of $A$ and $B$.  

We now impose some restrictions on the probabilistic models under consideration. First, we require all state spaces 
to be finite-dimensional (we are, after all, only 
attempting to recover {\em finite-dimensional} QM). Secondly, we require that models be {\em uniform}, in the 
sense that 
\begin{itemize} 
\item[(i)] all basic measurements have a common, finite number of outcomes, $n$, called the {\em rank} of $A$; and 
\item[(ii)] there exists a {\em maximally mixed} state, $\rho$, defined by $\rho(x) = 1/n$ for 
all basic measurement outcomes $x$ 
\end{itemize} 
These conditions are satisfied by finite-dimensional quantum-mechanical models, including those 
involving superselection sectors, provided that we restrict attention to {\em maximal} observables, i.e., 
those consisting of rank-one projections. More generally, condition (i) is reasonable if we think of basic measurements as {\em maximally informative}, so that each has the largest possible number of outcomes, and cannot be further refined.  
Given condition (i), 
the maximally-mixed state is well-defined mathematically, so in (ii), we are only requiring that it count as a 
physically accessible state. 

The following is a direct translation of the correlational features discussed above for quantum-mechanical systems, into the language of probabilistic models.

\begin{definition} 
A {\em conjugate} of a (uniform) probabilistic model $A$ is a model $\bar{A}$, together 
with an isomorphism $\gamma$ taking each basic measurement outcome $x$ of $A$ to an outcome $\bar{x} := \gamma(x)$ of $\bar{A}$, such that 
\begin{mlist} 
\item[(a)] Every state $\alpha$ of $A$ is the marginal of {\blue some} state  $\omega$ on $A$ and $\bar{A}$ 
({\blue in general, depending on $\alpha$}) correlating {\em some} basic measurement $E$ of $A$ with its counterpart on $\bar{A}$ so that for all $x \in E$, 
\[\omega(x,\bar{x}) = \alpha(x)\]
{\blue so that $\omega(x,\bar{y}) = \alpha(x) \delta_{x,y}$.} %for all outcomes $x$ of $E$.  Moreover, we require that 
\item[(b)] The maximally mixed state $\rho$ arises as the 
marginal of a bipartite state $\eta_A$ {\em uniformly} correlating {\em every} basic measurement with its counterpart, in the sense that 
\[\eta_A (x,\bar{x}) = \tfrac{1}{n}\]
for all basic measurement outcomes $x$, where $n$ is the rank of $A$. %number of outcomes in a basic measurement of $A$.
\end{mlist}
\end{definition} 

Evidently, in the quantum-mechanical case, where $\alpha$ corresponds to a density operator $W$, 
the state $\Psi_W$ supplies the correlating state $\omega$, while the bipartite state $\eta$ corresponds 
to the EPR state $\Psi_{\tfrac{1}{n}\1} = \Psi$. 

Mathematically, the existence of a conjugate system 
has affinities with the {\em purification postulate} of \cite{CDP}, 
though we do not require the correlating bipartite state $\omega$ above to be pure. 
Physically, a conjugate system $\bar{A}$ allows for the formation of records of the outcomes of measurements on $A$ 
in causally separated systems, exactly as in the quantum case.  
\tempout{To dilate on this, suppose we wish to perform one of the basic tests $E \in \M(A)$, 
while system $A$ is in state $\alpha$, and to leave a record of this state in some other system $B$, without 
any signal being sent from $A$ to $B$. Classically, this can be arranged by suitably correlating $A$ and $B$ 
in advance of measurement. In quantum mechanics, as observed above, it can be done by putting $A$ and $B$ into a suitable entangled state. For this to work for arbitrary probabilistic models $A$ and $B$, we would need the joint system $AB$ to be in some bipartite (and ``non-signaling") state $\omega$ with $A$-marginal $\alpha$. After measuring $E$ and obtaining an outcome, say $x$, $B$ is left in a conditional state $\omega_{2|x}$. For this to function as a record of the outcome $x$, it must be possible, at a later time, to {\em read off} which outcome occurred. Thus, there must be a test $F \in \M(B)$ that will allow us to discriminate sharply among the 
states $\{\omega_{2|x} | x \in E\}$. Thus, for each $x \in E$, there must be an outcome $y_x \in F$ 
with $\omega_{2|x}(y_x) = 1$, and with $y_x \not = y_{x'}$ for distinct outcomes $x, x' \in E$. 
Equivalently, there must be an injection $f : E \rightarrow F$ such that $\omega(x,y) = 0$ for $y \in F$ 
with $y \not = f(x)$.}
Condition (a) above simply requires that, for every state $\alpha$, there be 
at least one basic measurement on $A$ that can be thus recorded and later ``read off" by performing the corresponding measurement on $\bar{A}$. Condition (b) requires that, where $A$ is in the maximally mixed state, 
it be possible to record {\em every} basic measurement in this way. \\

\tempout{ 
\noindent{\em Remark:} One can simplify both the definition and the iterpretation of a conjugate by adding a bit 
more structure to the definition of probabilistic model. Specifically, suppose that any model $A$ carries 
a preferred group $G(A)$ of physical symmetries, acting alike on the set of basic measurements (and their outcomes), 
and on the set of states, in such a way that, for $g \in G(A)$, $(g\alpha)(x) = \alpha(g^{-1} x)$ for all 
states $\alpha$ and outcomes $x$. Suppose $G(A)$ is compact, and acts {\em transitively} on the set of basic measurements 
and on the set of basic measurement outcomes.  This guarantees uniformity, since we can take $\rho$ to be the invariant state obtained by group averaging. Adapting the foregoing definition to this setting, we can define an action of $G(A)$ on $\bar{A}$ by setting, e.g., $\bar{g}\bar{x} := \bar{gx}$ for basic measurement outcomes $x$. 
If we then add to condition (i) in the foregoing definition, the requirement that $\omega$ be invariant 
under physical symmetries leaving the state $\alpha$ fixed --- so that $\omega(g x, \bar{g} \bar{y}) = \omega(x,\bar{y})$ 
for all such symmetries $g$ --- then (ii) follows, since every $g \in G$ fixes the maximally mixed state $\rho$.
However, nothing in the derivation to follow depends 
in any way on, nor would it be simplified by, the imposition of such extra structure.\\ 
}

\noindent{\bf From correlation to Jordan algebras} 
Remarkably little is required, beyond the existence of a  conjugate, 
to secure a representation of $A$ in terms of a formally real Jordan algebra. This depends on a classic mathematical result, the {\em Koecher-Vinberg Theorem} \cite{FK}. 
A finite-dimensional ordered vector space $\Vec$ with positive cone $\Vec_+$ is {\em self-dual} if it carries an inner product such that $a \in \Vec_+$ iff $\langle a, b \rangle \geq 0$ for all $b \in \Vec_+$. 
If the group of invertible linear mappings $\Vec \rightarrow \Vec$ carrying $\Vec_+$ onto itself acts transitively on the {\em interior} of the cone $\Vec_+$, then $\Vec$ is said to be {\em homogeneous}. 
The Koecher-Vinberg Theorem asserts that if $\Vec$ is both homogeneous and self-dual, it can be endowed with 
a formally real Jordan structure for which $\Vec_+ = \{ a^2 | a \in \Vec\}$.  

Any probabilistic model $A$ gives rise in a natural way to two ordered vector spaces:  
a space $\V(A)$, generated by $A$'s states, and a space $\E(A) \leq \V(A)^{\ast}$ generated 
by evaluation functionals $\hat{x}: \alpha \mapsto \alpha(x)$ associated with basic measurement outcomes $x$.  
Since we are assuming that the state space is finite dimensional, both of the spaces $\E(A)$ and $\V(A)$ are also 
finite dimensional, and it is easy to see that they have the same dimension.
%The space $\E(A)$ also contains a {\em unit} functional, $u_A$, defined by $u_A(\alpha) = 1$ for every state $\alpha$. 
If we can show that $\E(A)$ is homogeneous and self-dual, then the Koecher-Vinberg 
Theorem will provide a formally real (equivalently, euclidean) Jordan structure on $\E(A)$ for which 
the cone of squares coincides with $\E(A)_+$. %for which $u_A$ is the Jordan unit. 

Call a probabilistic model $A$ {\em sharp} if, for every basic measurement outcome $x$, there is 
a {\em unique} state $\delta_{x}$ with $\delta_{x}(x) = 1$.  Physically, this is a way of saying that basic measurements are maximally informative: if we can predict the outcome with certainty, we know the system's state exactly.
%\\ %(a feature of both classical and quantum probability theory). \\

\begin{theorem} Suppose $A$ is sharp and has a conjugate. Then the state $\eta_{A}$ gives rise to a self-dualizing inner product on $\E(A)$, with respect to which $\E(A)$ and $\V(A)$ are isomorphic as ordered vector spaces.
\end{theorem} 

It follows that if $\V(A)$ is homogeneous, so is $\E(A)$, whence, by the Koecher-Vinberg Theorem, the latter carries a %canonical 
 formally real Jordan structure. 
But the homogeneity of $\V(A)$ has a direct physical interpretation: 
it says that for every {\em  non-singular} state --- that is, every state $\alpha$ with $\alpha(x) > 0$ for every basic measurement outcome --- there exists a {\em probabilistically reversible process $T$} --- {\blue defined below, 
but, roughly, one that can be reversed by another process with non-zero probability ---   such that} 
$T(\rho) = r \alpha$ {\blue where} $r \in [0,1]$. %= T(\rho)(u)$. {\blue [DEFINE $u$!]}  
In other words, every non-singular state can be {\em prepared}, up to normalization, by applying a probabilistically reversible process to the maximally mixed state. 

In fact, it is enough to assume less. 
By a {\em filter} for a basic measurement with outcome-set $E$, I mean a process $\Phi$ --- that is, a positive linear mapping $\Phi : \V(A) \rightarrow \V(A)$ --- that independently attenuates the reliability of each outcome $x \in E$, so that for every state $\alpha$, 
$\Phi(\alpha)(x) = t_x \alpha(x)$ for some constant $t_x$ (independent of $\alpha$). 
If we think of basic measurement outcomes as detectors, the existence of such filters, with arbitrary coefficients, is plausible; in standard QM, not only do they exist but, if $t_x > 0$ for every $x \in E$, then $\Phi$ can be chosen to be {\blue probabilistically} reversible. Where a probabilistic model $A$ shares this feature, I will say that $A$ {\em has arbitrary reversible filters}. 

\begin{corollary} Suppose that $A$ is sharp, has a conjugate, and has arbitrary reversible filters. Then $\E(A)$ is homogeneous and self-dual.
\end{corollary}

If we adopt a stronger assumption about filters, 
we can weaken the requirement that $A$ have a conjugate, and eliminate entirely the hypothesis that $A$ is sharp. 
Let us say that $A$ has a {\em weak} conjugate if the maximally mixed state $\rho$ is the marginal of a 
uniformly correlating state $\eta_A$ on $A$ and $\bar{A}$, as in condition (b) in Definition 1, but {\em not} assuming that {\em every} state is the marginal of a correlating state, i.e., not assuming condition (a).  That is, 
we require only that there exist a joint state on two copies of $A$ --- an analogue of the EPR state --- in which, if the same measurement is performed on each copy, the results are guaranteed to be the same, but are otherwise completely random. 

By applying the filter $\Phi$ to the system $A$, and then computing the canonical bipartite state $\eta_A$, we obtain a new bipartite state. Equally, we could begin by applying the counterpart of $\Phi$ to the conjugate system, obtaining another bipartite state. If these two bipartite states are in fact the same, we say that $\Phi$ is {\em symmetric}. %\\

\begin{corollary} 
Let $A$ have a weak conjugate. %system $\bar{A}$. I
If every non-singular state of $A$ can be prepared, up to normalization, from the maximally mixed state by a symmetric, reversible filter,  
then $\E(A)$ is homogeneous and self-dual. 
%{\blue [$\E(A)$ or $\E(A)_+$? See Cor. 1. Need to be consistent about this usage!]}
\end{corollary}

The proofs of these results are all quite short and straightforward. In summary, we recover euclidean Jordan algebras from either of two distinct, but related, sets of assumptions about 
physical systems represented by uniform, finite-dimensional probabilistic models:
\begin{itemize} 
\item[(a)] systems are sharp;
\item[(b)] Systems have conjugates; and 
\item[(c)] Systems have arbitrary reversible filters. 
\end{itemize} 
Alternatively, and more compactly: 
\begin{itemize} 
\item[(b$'$)] Systems have {\em weak} conjugates; 
\item[(c$'$)] All non-singular states can be prepared by reversible symmetric filters
\end{itemize}
Any euclidean Jordan algebra gives rise to a probabilistic model in which basic measurements correspond to 
{\blue \em Jordan frames}, i.e., sets $\{e_i\}$ of minimal idempotents satisfying $e_i \dot e_j= 0$ for $i \not = j$, and 
$\sum_{i} e_i = \1$, where $\1$ is the Jordan unit. {\blue (In standard QM, these would correspond to maximally fine-grained 
projective measurements.)}  In Appendix A, it is shown that any such model 
satisfies all of the assumptions above and, conversely, if $A$ satisfies 
either package of assumptions, the Jordan product on $\E(A)$ can be chosen so that {\blue the set 
of basic measurements} % $\M(A)$ 
is precisely the set of Jordan 
frames. 
%(and the {\blue unit effect} is the Jordan unit). 
Thus, these two sets of assumptions are in fact equivalent, and exactly characterize this class of euclidean Jordan-algebraic probabilistic models. 

There is actually a third possibility, in which condition 
(b) in the definition of a conjugate is replaced by a rather weak symmetry assumption and a 
version of Hardy's subspace axiom \cite{Hardy}. Again, the resulting package of assumptions is satisfied by, 
and hence, characterizes, Jordan models.  The details are spelled out in Appendix B.   
\\

\tempout{
{\bf The Subspace Axiom} In his pioneering paper \cite{Hardy}, Lucien Hardy made use of a powerful {\em subspace axiom}, 
requiring, in effect, that for every basic measurement outcome $x$, the set of states $\alpha \in \Omega(A)$ 
with $\alpha(x) = 0$ is itself effectively the state space of a ``subsystem", subject to all of the remaining 
axioms one wishes to impose. The later reconstructions \cite{Dakic-Brukner, Masanes-Mueller, CDP} all follow 
Hardy's lead in making use of one or another version of this postulate. It is noteworthy that Corollaries 1 and 2 
above do not depend on any such assumption.  However, if one finds the subspace axiom compelling, it can be 
put to good use in the present setting.  In Appendix B, it is shown that a subspace axiom slightly stronger than Hardy's, 
but having the same general character and motivation, together with a very weak and plausbible symmetry 
condition, can replace condition (b) in the definition of a conjugate.
}

\tempout{All of these assumptions can be expressed in less technical terms, albeit with some loss of precision. 
Condition (a) (sharpness) tells us that tests are maximally informative.  Condition (b) (existence of a conjugate) is really the combination of two somewhat similar principles, both of which can be expressed in ``Alice and Bob" language, as follows:
\begin{itemize}
\item[(b1)]  There is a state in which, if Alice and Bob peform the same test, they are certain to get the same outcome, but 
are totally uncertain as to which this will be;
\item[(b2)] For every state of Alice's system, there is a joint state and a measurement on Alice's system such that if Bob performs the same measurement, she and he will obtain the same result, {\em and} Alice's system is in the prescribed state.
\end{itemize} 
(As discussed earlier, (b$''$) can also be expressed in terms of there being, for every state on Alice's system, a measurement that can be recorded on Bob's system, without 
Alice having to send any data to Bob.) Finally, condition (c) is a reasonable methodological principle, requiring that a data-procesing task that we can certainly carry out classically  (that is, implementing a reversible filter with arbitrary nonzero coefficients) can be done in a physically reversible way.

In the second package, all this is paired down to just two assumptions: the very simple condition (b$'$), and a stronger version of (c), in which we insist that, not only can we implement 
arbitrary filters reversibly, but we can do so symmetrically with respect to Alice and Bob, and, what's more, prepare arbitrary states for Alice's system (or Bob's) in this way.  

It should be stressed that these are, in the language of \cite{BMU}, ``single-system" postulates, in that they apply to individual systems, rather than to an entire theory's worth of systems. In contrast, the subspace postulate 
requires that every subspace (roughly: face) of a state space be the state space of another system, {\em to which 
all other axioms continue to apply}. 
}

\noindent{\bf Other reconstructions of QM}  The approach of this paper offers some significant advantages over the reconstructions of quantum mechanics cited earlier. First, it is simply {\em easier}, in the sense that 
our results are obtained with less mathematical effort. (This, notwithstanding the length of this paper, 
which owes to the inclusion of many details intended to make the paper easier to follow.)
\footnote{And also notwithstanding my appeal to the  Koecher-Vinberg Theorem, as this is itself a very accessible result. See \cite{FK} for a not terribly taxing proof. A number of the other reconstructions mentioned here also depend on  nontrivial mathematical results, e.g., the classification of transitive actions of compact groups on spheres 
is used in \cite{Masanes-Mueller}.} 

Secondly, it  rests on fewer and (arguably) simpler assumptions. Certainly, the second package of asumptions (that is, (b$'$) and (c$'$) above) is smaller than anything found in earlier reconstructions of QM. 
Other reconstructions tend to impose strong constraints on subsystems, 
in effect assuming that every face of the state space corresponds to the state space of a ``sub-system", satisfying the remaining axioms. Nothing like this is needed here (though, as mentioned above, if one finds a such a ``subspace axiom"
compelling, it can be put to good use in the present approach; see Appendix B for details).  
A related assumption, also used in several of the cited papers, is that all systems having the same ``information capacity" --- the maximal number of states sharply distinguishable by single-shot measurement --- are isomorphic. 
The present approach entirely avoids such an assumption. It also does without  
the assumption, commonly called the {\em no-restriction hypothesis} \cite{Janotta-Lal}, used in \cite{Masanes-Mueller} for bits, that all mathematically possible {\blue \em effects}  --- {\blue that is, affine functionals assigning probabilities to states} --- correspond to physically accessible measurement results.  
More recently, the interesting paper \cite{BMU} derives the same Jordan-algebraic structure arrived at here, 
but in a different way. In addition to a strong symmetry postulate, this paper assumes a weak form of the no restriction 
hypothesis, namely, that all finite sets of ``allowed" effects that sum to the {\blue \em unit effect (the effect 
identically 1 on all states)} correspond to accessible measurements, along with a kind of spectral decomposition for states. Here, we manage without any form of no-restriction assumption, and a spectral decomposition for states is {\em derived}, rather than postulated. 

Finally, %and more importantly, the present approach is {\em more flexible}. 
all of the earlier reconstructions of QM cited above assume some form of {\blue {\em local tomography}. This is the doctrine that the state of a bipartite system is determined by the joint probabilities 
it assigns to outcomes of  measurements on the two component systems. } This principle has a certain intuitive appeal; moreover, it is 
well known, and easy to see on dimensional grounds, that among finite-dimensional real, complex and quaternionic quantum mechanics, only in the complex version are composites locally tomographic. 

More generally \cite{Hanche-Olsen, LTHSD}, the only probabilistic theory in which systems correspond to Jordan models and composite systems are locally tomographic, and which includes at least one system having the structure of a qubit, is finite-dimensional complex quantum  mechanics.  Thus, if one insists on local tomography, it can be added to the list of assumptions discussed above, 
and leads to standard, complex QM (with superselection rules). 
%All of the reconstructions of QM cited above assume some form of local tomography. 
One should perhaps not rush to embrace local tomography as a universal principle, however. The very fact that it excludes real and quaternionic quantum theory suggests that it is too strong. {\blue 
%This is 
%both because 
%there is a sense in which real and complex Hilbert spaces can be interpreted 
%$J^2 = \1$ in the real case, and $J^2 = -\1$ in the quaternionic,  
%For one thing,
% both because 
There are natural ways of representing complex Hilbert spaces in terms of real or quaternionic ones, 
and vice versa\footnote{A real or quaternionic Hilbert space can be regarded as a complex Hilbert space 
equipped with a designated anti-unitary operator $J$ satisfying, respectively, $J^2 = \1$ or $J^2 = -\1$; 
conversely, a complex Hilbert space is essentially equivalent to a real or quaternionic one equipped with, 
respectively, an orthogonal or a simplectic operator $J$ satisfying $J^2 = -\1$ or $J^2 = \1$.};
%quantum system endowed with extra structure (respectively, a designated orthogonal or symplectic transformation $J$ %with  $\J^2 = -\1$), 
%and also because  
moreover, these representations have physical meaning, in that bosonic or fermionic (complex) quantum systems can very naturally be modelled in terms of the corresponding real or, respectively, quaternionic Hilbert spaces.  Again, see, e.g., \cite{Baez} a cogent development of this line of thought.} 
%for some cogent reasons {\em not} to exclude these cases. 
In any case,  it seems valuable to be able to delineate clearly what does and what does not depend on this assumption, particularly if we are interested in the possibilities for a ``post-quantum" theory.
\\

\noindent{\bf Organization} The balance of this paper is arranged as follows. Section 2 provides general background on probabilistic models, ordered vector spaces, Jordan algebas and so on, making more precise many of the technical terms used above.  This material will be familiar to many, but probably not to all, readers. Section 3 contains the proof of Theorem 1; Corollaries 1 and 2 are proved in Section 4. Section 5 collects some final thoughts, inlcuding a few further remarks on how the approach of this paper compares to the reconstructions of QM cited above. Appendix A contains additional information on probabilistic models associated with euclidean Jordan algebras, and Appendix B shows how a version of the ``subspace axiom", plus a symmetry assumption, can replace some of the assumptions native to this paper. 

Several of the ideas developed here were earlier explored, and somewhat similar results derived, in  \cite{Wilce09} and \cite{SSD}, but the approach taken here is much simpler and more direct, and seems to go a good deal farther. 

\section{Background}

The mathematical framework for this paper is that of ``generalized probabilistic theories" \cite{Barrett}, 
in the idiom of \cite{Wilce09, BW12}, 
which I now quickly review.\footnote{This is a variant of the standard ``convex-operational" framework 
developed in the 1960s and 1970s by Ludwig, Davies and Lewis and others (e.g., \cite{Ludwig, Davies-Lewis, Edwards}),  specialized to finite dimensions, and with additional structure deriving from work of Foulis and Randall \cite{FR}.}
In a few places, set off in numbered definitions, my usage differs slightly from that of these last-cited works. 
 See \cite{FK, Alfsen-Shultz} for more information on ordered vector spaces and Jordan algebras. \\

\noindent{\bf Ordered vector spaces} 
An {\em ordered vector space} is a real vector space $\Vec$ equipped with a closed, convex cone $\Vec_+$ with $\Vec_+ \cap -\Vec_+ = \{0\}$ and $\Vec = \Vec_+ - \Vec_+$ --- that is, $\Vec$ is spanned by $\Vec_+$. The cone induces a partial 
order, invariant under translation and multiplication by non-negative scalars, given by $a \leq b$ iff $b - a \in \Vec_+$. 
As an illustration, the space $\R^{X}$ of real-valued functions on a set $X$ is ordered by the cone $\R^{X}_+$ of functions taking non-negative values. Another example is the space $\L_{\sa}(\H)$ of hermitian operators on a real, complex or quaternionic Hilbert space, ordered by the cone of positive semi-definite operators. 

A linear mapping $T : \Vec \rightarrow \F$ between ordered vector spaces is {\em positive} iff $T(\Vec_+) \subseteq \F_+$. If $T$ is bijective and $T^{-1}$ is also positive, then 
$T$ is an {\em order isomorphism}. If $\Vec$ and $\F$ are finite dimensional with $\dim(\Vec) = \dim(\F)$, $T$ is an order isomorphism iff $T(\Vec_+) = \F_+$. 
The dual space $\Vec^{\ast}$ of a finite-dimensional ordered linear space carries a natural ordering, defined 
by the  {\em dual cone}, $\Vec^{\ast}_+$, consisting of positive linear functionals $f \in \Vec^{\ast}$. 
\\

\noindent{\bf Probabilistic Models} As discussed above, a {\em probabilistic model} is characterized by a set $\M(A)$ of basic measurements or {\em tests}, and a set $\Omega(A)$ of states. It is convenient to identify each test with 
its outcome-set, so that $\M(A)$ is simply a collection of non-empty sets (a {\em test space}, in the language of \cite{BW12}). Let $X(A)$ stand for the union of this collection; that is, $X(A)$ is the space of all outcomes of all basic measurements. States are understood as assignments of probabilities to measurement-outcomes, that is, as functions $\alpha : X(A) \rightarrow [0,1]$ such that $\sum_{x \in E} \alpha(x) = 1$ for all tests $E \in \M(A)$ (but not 
all such functions necessarily correspond to states).  As mentioned 
above, a state $\alpha \in \Omega(A)$ is {\em non-singular} iff $\alpha(x) > 0$ for all $x \in X(A)$. 
To reflect the possibility of forming statistical mixtures, I also assume that $\Omega(A)$ is convex, 
that is, if $p_1,...,p_n$ are non-negative real numbers summing to $1$, and $\alpha_1,...,\alpha_n$ are 
states in $\Omega(A)$, then the function $p_1 \alpha_1 + \cdots + p_n \alpha_n$ also belongs to $\Omega(A)$. 
Finally, I assume that $\Omega(A)$ is closed under pointwise limits, whence, compact as a subset of 
$[0,1]^{X(A)}$ in the product topology.

By way of illustration, in the simplest {\em classical} model, $\M(A)$ consists of a single, finite test, and $\Omega(A)$ is the simplex of all probability  weights on that test. 
Of more immediate interest to us is the {\em quantum model} $A(\H) = (\M(\H),\Omega(\H))$ associated with a complex Hilbert space $\H$. The test space $\M(\H)$ is the set of orthonormal bases of $\H$; thus, the outcome-space $X(\H)$ is the set of unit vectors of $\H$. The state space $\Omega(\H)$ consists of the quadratic forms associated with density operators on $\H$, so that a state $\alpha \in \Omega(\H)$ has the form $\alpha(x) = \langle W_{\alpha} x, x \rangle$ for some density operator $W_{\alpha}$, and all unit vectors $x \in X(\H)$. Real and quaternionic 
quantum models, corresponding to real or quaternionic Hilbert spaces, are defined in the same way.\\

\noindent
{\em Remark:}  Not {\em every} physically accessible observable on a finite-dimensional quantum system is represented by an orthonormal basis. Rather, the general observable corresponds to a positive-operator-valued measure. Similarly, for an arbitrary probabilistic model $A$, the test space $\M(A)$ {\em may}, but need not, represent a complete catalogue of all possible measurements one might make on the system represented by $A$: rather, it is some privileged (or perhaps,  
simply {\em convenient}) catalogue of such measurements, sufficiently large to determine the system's states. 
\\

\noindent{\bf The spaces $\V(A)$ and $\E(A)$.} 
Any probabilistic model $A$ gives rise in a canonical way to an ordered vector space $\V(A)$. This is simply the 
span of the state space $\Omega(A)$ in the space $\R^{X(A)}$, ordered by the cone $\V(A)_+$ of non-negative multiples of states; that is, $\beta \in \V(A)_+$ iff $\beta = t \alpha$ for some {\blue state $\alpha \in \Omega(A)$ and 
some real} constant $t \geq 0$.  
An element of $\V(A)_+$ of the form $t\alpha$ with $\alpha \in \Omega(A)$ and $t \leq 1$ is said to be {\em sub-normalized}.  One can show that the interior of $\V(A)_+$ consists exactly of multiples of non-singular states. 
The {\em dimension} of a model $A$ is the dimension of $\V(A)$. As mentioned in the introduction,
 {\em  it is assumed in this paper that all probabilistic models are finite-dimensional.}

{\blue There is a canonical positive linear functional $u_A : \V(A) \rightarrow \R$, called the 
{\em unit effect} of $A$, given by 
$u_{A}(\alpha) = \sum_{x \in E} \alpha(x)$, where $E$ is any test in $\M(A)$. Note that if $\alpha \in \V(A)_+$, then 
$u_{A}(\alpha) = 1$ precisely when $\alpha \in \Omega(A)$, and that every non-zero element $\alpha \in \V(A)_+$ 
has the form $\beta = t \alpha$ for a unique $t = u_A(\beta) > 0$. Thus, any non-zero $\beta \in \V(A)_+$ 
can be {\em normalized} to yield a state $\tilde{\beta} = u_{A}(\beta)^{-1} \beta \in \Omega(A)$. }
A positive linear functional $f \in \V(A)^{\ast}_{+}$ with $f \leq {\blue u_A}$  is usually called an {\em effect} on $A$, and can be thought of as representing 
an ``in-principle" measurement outcome, with probability $0 \leq f(\alpha) \leq 1$ in state $\alpha$. 
Every outcome $x \in X(A)$ corresponds to an effect  $\hat{x} : \V(A) \rightarrow \R$, given by $\hat{x}(\alpha) = \alpha(x)$ for all $\alpha \in \V(A)$, {\blue and $u_A = \sum_{x \in E} \hat{x}$.}%Hence, we have a natural injection $X(A) \rightarrow \V(A)^{\ast}$. 
\tempout{We define the {\em unit effect} $u_A \in \V(A)^{\ast}$ by 
$u_A := \sum_{x \in E} \hat{x}$, where $E$ is any test in $\M(A)$. This is independent of $E$, 
since $u_A(\alpha) = 1$ for all $\alpha \in \Omega(A)$. More generally, if $\beta \in \V(A)_+$, say $\beta = t\alpha$, $t \geq 0$, 
then $u_A(\beta) = t$.  Thus, any {\blue non-zero} $\beta \in \V(A)_+$ can be {\em normalized} to 
yield a state $\tilde{\beta} := u_{A}(\beta)^{-1} \beta \in \Omega(A)$. } 

If $A = A(\H)$ is the quantum mechanical model associated with a finite-dimensional Hilbert space $\H$, as 
discussed above, then, identifying $\Omega(\H)$ with the convex set of density operators on $\H$, 
$\V(A)$ can be identified with the ordered vector space $\L(\H)$ of self-adjoint operators on $\H$, with its 
usual cone {\blue of positive operators}.  We can then also identify $\V(\H)^{\ast}$ with $\L(\H)$, using the 
trace inner product. That is, if $a \in \L(\H)$, we can define a positive linear functional $a \in \V(A)^{\ast}$ by  setting 
$a(\alpha) = \Tr(a\alpha)$ for all $\alpha \in \V(A) = \L(\H)$, and all such functionals arise uniquely from elements of $\L(\H)$. 
In this setting, measurement-outcomes are unit vectors of $\H$, and, 
for each $x \in X(\H)$, and for each density operator $\alpha$ on $\V(A)$, $\hat{x}(\alpha) = \Tr(\alpha p_x)$ where $p_x$ is the rank-one projection associated with $x$. More generally, effects 
correspond to positive operators between $\0$ and $\1$. 

The spectral theorem for self-adjoint operators tells us that every effect in $\V(A(\H)) = \L(\H)$ is a positive linear combination of functionals $\hat{x}$ corresponding to measurement outcomes. This will not be 
the case for probabilistic models in general. 
 It is therefore useful to define a smaller cone, as follows: 

\begin{definition} 
 The space $\E(A)$ is the span, in $\V(A)^{\ast}$, of the set of effects $\hat{x}$ associated with 
outcomes $x \in X(A)$, ordered by the cone $\E(A)_+$ of finite linear combinations 
 $\sum_{i} t_i \hat{x}_i$, $x_i \in X(A)$, with coefficients $t_i \geq 0$.
\end{definition} 

Since we are assuming that $\V(A)$ and (hence) $\V(A)^{\ast}$ are finite-dimensional, the set of functionals $\hat{x}$ in fact spans $\V(A)^{\ast}$ (since 
they separate points of $\V(A)$.)  Thus, $\E(A)$ and $\V(A)^{\ast}$ are identical {\em as vector spaces}. However, their cones are generally quite different. 
If $a \in \E(A)_+$, then $a(\alpha) \geq 0$ for all $\alpha \in \V(A)_+$, so the cone $\E(A)_+$ is 
 contained in the dual cone $\V(A)^{\ast}_+$, but the inclusion is usually proper.  Thus, $\E(A)$ and $\V(A)$ are generally distinct as {\em ordered} vector spaces.\\
%\footnote{The pair $(\V(A),\E(A))$ is a finite-dimensional 
%instance of a {\em base-normed and order-unit space in separating statistical duality}, in the language 
%of \cite{Davies-Lewis} [?]}\\

\noindent{\em Remark:} 
The space $\E(A)$ will be a useful technical tool in what follows, but should not necessarily be regarded as 
anything more than that. In particular, it is {\em not} assumed that all physically meaningful effects reside 
in $\E(A)_+$, nor that every effect in $\E(A)_+$ is physically meaningful. In fact, it will not be necessary 
to take any position at all on which effects, other than those associated with measurement outcomes, are physically significant. %\footnote{
Thus, as mentioned in the introduction, we avoid the 
so-called {\em no-restriction hypothesis}  \cite{Janotta-Lal}, {\blue namely, the asumption that} {\em all} effects in $\V(A)^{\ast}_+$ are physically accessible.  This assumption is often made in the literature, sometimes 
explicitly (e.g., \cite{Masanes-Mueller}), sometimes not.\\
%}\\

\noindent{\bf Processes}  
A physical process on a system represented by a probabilistic model $A$ is naturally represented by an affine (that is, convex-linear) mapping $T : \Omega(A) \rightarrow \V(A)$ such that, for every $\alpha \in \Omega(A)$, $T(\alpha) = p \beta$ for some $\beta \in \Omega(A)$ and some constant $0 \leq p \leq 1$ (depending on $\alpha$), which we can regard 
as the probability that the process occurs, given that the initial state is $\alpha$.\footnote{To be 
clear, we are not suggesting that {\em all} such positive mappings on $\V(A)$ represent physically allowable processes. Indeed, in QM, only completely positive mappings are physically allowable.}  
%If $p = 1$, we may say the process is {\em lossless}. 
Such a mapping extends 
uniquely to a  positive linear mapping $T : \V(A) \rightarrow \V(A)$ with $T(\alpha)(u_A) \leq 1$ for all $\alpha \in \Omega(A)$. 
%{\bf Say that a process $T : \V(A) \rightarrow \V(B)$ {\em prepares} a state $\alpha$ of $A$ from another state, $\beta$, %if $\alpha$ is a multiple of $T(\beta)$, i.e., $T(\beta)$ coincides with $\alpha$ up to normalization. 
%[below?]}

\begin{definition}  A process $T : \V(A) \rightarrow \V(A)$ is {\em probabilistically reversible} --- hereafter, just {\em p-reversible}\footnote{It should be noted that my usage is slightly nonstandard here: ordinarily, the adjective {\em reversible} is reserved for processes that are probabilistically reversible, in the above sense, with probability one.} --- iff there is another process, $S$, such that, for every state $\alpha$, there exists a constant $p \in (0,1]$ with $S(T(\alpha)) = p \alpha$. 
\end{definition} 

In other words, $S$ allows us to recover $\alpha$ from $T(\alpha)$, {\em up to normalization}. It is not hard to see that $p$ must be independent of $\alpha$, so that $S = p T^{-1}$. In particular, $T$ is an order-automorphism of $\V(A)$.

A process $T : \V(A) \rightarrow \V(B)$ has a dual action on $V(A)^{\ast}$, given by $T^{\ast}(f) = f \circ T$ for all $f \in \V(A)^{\ast}$, with $T^{\ast}(u_A) \leq u_A$. $T$ is lossless iff $T^{\ast}(u_A) = u_A$. In our finite-dimensional setting, we can identify $\V(A)^{\ast}$ with $\E(A)$ as vector spaces, but not, generally, as ordered vector spaces. While {\blue $T^{\ast}$} will 
preserve the dual cone $\V(A)^{\ast}_{+}$, it is {\em not} required, {\em a priori}, that 
$T^{\ast}$ preserve the cone $\E(A)_{+} \leq \V(A)^{\ast}$. This reflects the idea that not every physically accessible 
measurement need appear among the tests in $\M(A)$, as discussed above.\\

\noindent{\bf Self-Duality and Jordan Algebras.}  
For both classical and quantum models, the ordered spaces $\E(A)$ and $\V(A)$ are isomorphic. 
In the former case, where $\M(A)$ consists of a single test $E$ and $\Omega(A)$ is the simplex of all probability weights on $E$, we have $\V(A) \simeq \R^{E}$ and $\E(A) \simeq (\R^{E})^{\ast}$, with the standard inner product on $\R^{E}$ providing the order-isomorphism. 
If $\H$ is a finite-dimensional real or complex Hilbert space,  we have an affine isomorphism between the state space 
of $\Omega(\H)$ and the set of density operators on $\H$, allowing us to identify $\V(A(\H))$ with  {\red 
the space $\L_{\sa}(\H)$ of self-adjoint operators on $\H$, ordered by the cone of positive operators}. For any $x \in X(\H)$, the evaluation functional $\hat{x} \in \V(A)$ is then given by $W \mapsto \langle Wx, x \rangle = \Tr(WP_x)$. It follows that $\E(A(\H)) \simeq \L_{\sa}(\H)^{\ast} \simeq \L_{\sa}(\H)$, with the latter isomorphism implemented by the trace inner product.

More generally, call an inner product $\langle ~,~  \rangle$ on an ordered vector space $\Vec$ {\em positive} 
iff $\langle a, b \rangle \geq 0$ for all $a, b \in \Vec_+$. We then have a positive linear mapping $\Vec \rightarrow \Vec^{\ast}$, namely $a \mapsto \langle a, \cdot \rangle$. If this is an order-isomorphism, one says that $\Vec$ is {\em self-dual} with respect to this inner product. This is equivalent to the condition $a \in \Vec_+$ iff $\langle a, b \rangle \geq 0$ for all $b \in \Vec_+$. In this language, the standard inner product on $\R^{E}$ and the trace inner product on 
$\L_{\sa}(\H)$ are self-dualizing, for any finite set $E$ and finite-dimensional Hilbert space $\H$. 
 
In fact, any {\em euclidean Jordan algebra}, ordered by its cone of squares, is self-dual with respect to its canonical inner product. Recall here that a Jordan algebra is a real
commutative (but not necessarily associative) unital algebra $(\J,\dot)$ 
satisfying the {\em Jordan identity} 
\[a \cdot(a^2 \dot b) = a^2 \dot (a \dot b)\]
for all $a, b \in \J$ (with $a^2 = a \dot a$). A {\em euclidean}  Jordan algebra (EJA) is 
a finite-dimensional Jordan algebra $\J$ equipped with an inner product $\langle ~ , ~ \rangle$ such that 
$\langle a \dot b , c \rangle = \langle b ,  a \dot c \rangle$ 
for all $a, b, c \in \J$. This is equivalent to the condition that $\J$ be {\em formally real}, i.e, 
that $\sum_{i=1}^{k} a_{i}^{2} = 0$ implies $a_i = 0$ for all $i$. 
A EJA $\J$ is also an ordered vector space with 
positive cone $\J_+ = \{ a^2 | a \in \E\}$, and it can be shown that this cone 
is self-dual with respect to the given inner product \cite{FK}. 
Examples of euclidean Jordan algebras include the space ${\cal L}_{\sa}(\H)$ of self-adjoint operators on a 
finite-dimensional real, complex or quaternionic Hilbert space $\H$, with $a \cdot b = \frac{1}{2}(ab + ba)$, and with $\langle a | b \rangle = \Tr(ab)$. The {\em exceptional 
Jordan algebra} of self-adjoint 
hermitian matrices over the Octonions is also a euclidean Jordan algebra. Finally, one obtains a euclidean Jordan 
algebra, called a {\em spin factor}, by defining on $V_n := \R \times {\R}^{n}$ a product 
$(t,\x) \cdot (s,\y) = (ts + \langle \x, \y\rangle, t\y + s\x)$. 
This essentially exhausts the possibilities: according to the Jordan-von Neumann-Wigner classification theorem \cite{JNW}, every euclidean Jordan algebra is a direct sum of euclidean Jordan algebras of these five types.

We can associate a probabilistic model to an EJA $\J$ in the following way. An 
{\em idempotent} in $\J$ is an element $e = e^2$. An idempotent is {\em minimal}, or {\em primitive}, iff 
for any idempotent $f \leq e$, $f = 0$ or $f = e$. Two idempotents $e, f$ are {\em  Jordan-orthogonal} iff 
$e \dot f = 0$. A maximal pairwise Jordan-orthogonal set $\{e_1,...,e_n\}$ of primitive idempotents summing 
to the Jordan unit is 
called a {\em Jordan frame}. 

\begin{definition} 
The {\em Jordan model} $A(\J) = (X(\J),\M(\J),\Omega(\J))$  
corresponding to a euclidean Jordan algebra $\J$ has 
$X(\J)$ the set of primitive idempotents, $\M(\J)$, the set of Jordan frames of $\J$, and 
$\Omega(\J)$ the set of states of the form 
$\alpha(x) := \langle a, x \rangle$ 
where $a \in \E(A)_+$ satisfies $\langle a, \1 \rangle = 1$. The spectral theorem for EJAs (\cite{FK}, Theorem III.1.2) expresses every $a \in \J$ in the form 
$a = \sum_{x \in E} t_{x} x$ where $E$ is a Jordan frame. Therefore, $\J = \E(A)$, and the model 
$A(\J)$ is self-dual.
% $\1$ being the Jordan unit of $\J$. 
\end{definition}

Besides self-duality, all euclidean Jordan algebras share a property called {\em homogeneity}: the group of 
order-automorphisms of $\J$ acts transitively on the {\em interior} of the positive cone $\J_+$. 
The {\em Koecher-Vinberg Theorem} \cite{FK} states that, conversely, any finite-dimensional 
homogeneous, self-dual, homogenous ordered vector space $\J$ can be equipped with the structure of a euclidean Jordan algebra.

\begin{definition} 
A probabilistic model $A$ is {\em homogeneous} iff $\V(A)$ is homogeneous, and {\em self-dual} iff $\E(A)$ carries an 
inner product with respect to which it is self-dual {\em and} $\E(A)_+ \simeq \V(A)_+$, in the sense that 
$a \in \E(A)_+$ iff $\alpha(x) := \langle a, \hat{x} \rangle$ defines an element of $\V(A)_+$, and every 
element of $\V(A)_+$ arises in this way.
\end{definition} 

If $A$ is both homogeneous and self-dual --- henceforth, HSD --- then $\E(A)$ is also homogeneous and self-dual, and thus, by the Koecher-Vinberg Theorem, can be made into a Jordan algebra. In Appendix A, it is shown that this can be 
done in such a way that $A$ is actually isomorphic to the Jordan model corresponding to $\E(A)$.\\

\noindent{\bf Bipartite States and Conditioning}  %and Composite Systems} 
A {\em joint probability weight} on a pair of models $A$ and $B$ is a mapping $\omega : X(A) \times X(B) \rightarrow \R$ such that, 
for all $E \in \M(A)$ and $F \in \M(B)$, 
\[\sum_{(x,y) \in E \times F} \omega(x,y) = 1.\]
Such a weight is said to be {\em non-signaling} if, in addition, the {\em marginal weights} 
\[\omega_{1}(x) := \sum_{y \in F} \omega(x,y) \ \ \mbox{and} \ \ \omega_{2}(y) := \sum_{x \in E} \omega(x,y) \]
are well-defined, i.e., independent of the choice of tests $F \in \M(B)$ and $E \in \M(A)$, respectively. 
The idea is that such a state precludes the sending of signals between $A$ and $B$ based solely on the choice of what test to perform. 

If $\omega$ is non-signalling, 
then given outcomes $y \in X(B)$ and $x \in X(A)$, we can 
define {\em conditional} probability weights $\omega_{1|y}$ and $\omega_{2|x}$ on $A$ and $B$, respectively, by setting 
\[\omega_{1|y}(x) =  \frac{\omega(x,y)}{\omega_{2}(y)} \ \mbox{and} \ \omega_{2|x}(y) := \frac{\omega(x,y)}{\omega_{1}(x)},\]
when $\omega_{2}(y)$ and $\omega_{1}(x)$ are non-zero. This gives us the following bipartite {\em law of total probability} \cite{FR}
\begin{equation} \omega_{2} = \sum_{x \in E} \omega_1(x) \omega_{2|x} \ \ \mbox{and} \ \ \omega_1 = \sum_{y \in F} \omega_{2}(y) \omega_{1|y}\end{equation}
which will be exploited below. %\\

\begin{definition} Let $\omega$ be a non-signaling joint probability weight on $A$ and $B$. If all conditional weights $\omega_{1|x}$ and $\omega_{2|y}$ (and hence, the marginals $\omega_1$ and $\omega_2$) of $\omega$ belong to $\Omega(A)$ and $\Omega(B)$, respectively, then we say that 
$\omega$ is a {\em bipartite state} on the models $A$ and $B$. 
\end{definition} 
%\\
If $\H$ and $\K$ are real or complex Hilbert spaces, every density operator $W$ on $\H \otimes \K$ gives rise to a bipartite state on $A(\H)$ and $A(\K)$, given by $\omega(x,y) = \langle W x \otimes y, x \otimes y \rangle$. \\

\tempout{
For a simple example of a bipartite state that is neither classical nor quantum, let $B$ denote the model consisting of a 
test space $\B = \{\{x,y\}, \{a,b\}\}$ with two non-overlapping, two-outcome tests, and with $\Omega$ the set of all probability weights thereon. (Geometrically, the latter is simply the square 
$[0,1]^{\{x,a\}}$; accordingly, this model is sometimes called the {\em square bit}.) The bipartite state on $B \times B$ given by the table below (a variant of the ``non-signaling box" of
Popescu and Rohrlich \cite{PR}) is clearly non-signaling. Notice that it also establishes a perfect, uniform correlation between the outcomes of any test on the first system and its counterpart 
on the second.  
\[\begin{array}{|c|cc|cc|}
\hline
  &  x  &  y  &  a  &  b  \\
\hline
x & 1/2 &  0  & 1/2 &  0  \\ 
y & 0   & 1/2 &  0  & 1/2 \\ 
a & 0   & 1/2 & 1/2  & 0  \\ 
b & 1/2 &  0  &  0  & 1/2 \\
\hline
\end{array}\]
}

\noindent{\bf The conditioning map} If $\omega$ is a bipartite state on $A$ and $B$, define the associated 
{\em conditioning maps} $\hat{\omega}: X(A) \rightarrow \V(B)$ and $\hat{\omega}^{\ast} : X(B) \rightarrow \V(A)$ by 
\[\hat{\omega}(x)(y) \ = \ \omega(x,y)  \ = \ \hat{\omega}^{\ast}(y)(x).\]
Note that $\hat{\omega}(x) = \omega_1(x) \omega_{2|x}$ for every $x \in X(A)$, i.e., 
$\hat{\omega}(x)$ can be understood as the {\em un-normalized} conditional state of $B$ given the outcome 
$x$ on $A$, and similarly for $\hat{\omega}^{\ast}(y)$. 

The conditioning map $\hat{\omega}$ extends uniquely to a positive linear mapping $\E(A) \rightarrow \V(B)$, which I also denote 
by $\hat{\omega}$, such that $\hat{\omega}(\hat{x}) = \hat{\omega}(x)$ for all outcomes $x \in X(A)$. 
(To see this, consider the positive linear mapping $T : \V(A)^{\ast} \rightarrow \R^{X(B)}$ 
defined, for $f \in \V(A)^{\ast}$, by $T(f)(y) = f(\hat{\omega}^{\ast}(y))$ for all  $y \in X(B)$. 
If $f = \hat{x}$,  we have $T(\hat{x}) = \omega_{1}(x) \omega_{2|x} \in \V(B)_{+}$. Thus, 
the range of $T$ lies in $\V(B)$.) 
%and, moreover, $T$ is {\em positive} on the cone $\E(A)_+$. 
%Hence, $T$ defines a positive linear mapping $\E(B) \rightarrow \V(A)$, as advertised. 
In the same way, $\hat{\omega}^{\ast}$ defines a positive linear mapping $\hat{\omega}^{\ast} : \E(B) \rightarrow \V(A)$. 
Notice that $\hat{\omega}$ need {\em not} take $\V(A)^{\ast}_+$ into $\V(B)_+$. This is the principal reason 
for working with $\E(A)$ rather than $\V(A)^{\ast}$. {\blue If $\hat{\omega} : \E(A) \rightarrow \V(A)$ is an order isomorphism, $\omega$ is said to be an {\em isomorphism state} \cite{BGW}.} \\

\noindent{\bf Composite Systems}
As the language here suggests, one wants to view (some) bipartite states as elements of the state space of a {\em composite model}.  Broadly, a {\em composite} of two probabilistic models $A$ and $B$, is a model $AB$ equipped with a mapping 
$X(A) \times X(B) \rightarrow \V(AB)_{+}^{\ast}$, taking every pair of outcomes $x \in X(A)$ and $y \in X(B)$, to a product 
{\em effect} $xy$, such that every state $\omega \in \Omega(AB)$ pulls back to a bipartite state  
$\omega(x,y) := \omega(xy)$ on $A$ and $B$. While nothing in the mathematical development to follow depends on the choice of such a composite model, questions of interpretation may hinge on such a choice. \\

\section{Conjugate Systems}

Let $\H$ be an $n$-dimensional complex Hilbert space and $\bar{\H}$ is its conjugate space. As discussed in the introduction, the maximally engangled ``EPR" state, defined by 
$\Psi = \frac{1}{\blue \sqrt{n}}\sum_{x \in E} x \otimes \bar{x}$, 
where $E$ is any orthonormal basis for $\H$, establishes a perfect, uniform correlation betweeen every projection-valued observable on the system associated with $\H$, and its counterpart on 
$\bar{\H}$. Moreover, $\Psi$ effectively {\em defines} the normalized trace inner product on $\E(\H) = \L_{\sa}(\H)$. Since it is precisely this inner product that makes $\L_{\sa}(\H)$ self-dual, one might guess that the existence of a uniformly correlating bipartite state is implicated in self-duality  more generally.

As a first step, we need to generalize the relationship between the models $A(\H)$ and $A(\bar{\H})$. In order to do this, as mentioned in the introduction, 
we need first to impose the minor restriction that, henceforth, {\em all models are uniform}, meaning 
that all tests have a common cardinality $n$, and that the {\em maximally mixed} state $\rho$, given 
by $\rho(x) = 1/n$ for all $x \in X(A)$, belongs to $\Omega(A)$.  An {\em isomorphism} between two models $A$ and $B$ is a bijection $\phi : X(A) \rightarrow X(B)$ such that $\phi(E) \in \M(B)$ iff $E \in \M(A)$, and 
$\beta \circ \phi \in \Omega(A)$ iff $\beta \in \Omega(B)$. It is straightforward that such a mapping gives rise to an order isomorphism --- which I'll also denote by $\phi$ --- from $\E(A)$ to $\E(B)$, defined by $\phi(\hat{x}) = \hat{\phi(x)}$ for all $x \in X(A)$. 
The following reprises, and makes more precise, the definition of a conjugate system (Definition 1).
\\

\noindent{\bf Definition 1 (bis)} Let $A$ be a uniform probabilistic model of rank $n$. A {\em conjugate} for $A$ is a triple $(\bar{A},\gamma_A, \eta_A)$ 
consisting of a probabilistic model $\bar{A}$, 
an isomorphism $\gamma_A : A \simeq \bar{A}$, and a bipartite state $\eta_A$ on $A$ and $\bar{A}$ %\footnote{That is, on some composite $A\bar{A}$}, 
such that 
\begin{itemize}
\item[(a)] $\eta_{A}(x, \gamma_A(x)) = 1/n$ for 
every $x \in X(A)$. 
\item[(b)] Every state $\alpha \in \Omega(A)$ is the marginal of some bipartite state $\omega$ on $A$ and $\bar{A}$ that correlates {\em some} test $E \in \M(A)$ with its counterpart on $\bar{A}$, so that $\omega(x,\gamma(x)) = \alpha(x)$ for every $x \in E$. 
\end{itemize}  
{\blue As remarked earlier, the marginals of the perfectly correlating state $\eta$ in part (a) are the maximally mixed 
states $\rho$ on $A$ and $\bar{\rho}$ on $A$ and $\bar{A}$, respectively.}
Where no ambiguity is likely, I write $\bar{x}$ for $\gamma_{A}(x)$. 
If $\bar{A}$ satisfies (a), but not necessarily (b), I will call it a {\em weak conjugate} for $A$.\\
% (To emphasise this distinction, I will sometimes refer to conjugates as {\em strong} conjugates.)\\

Given any bipartite state on $A$ and $\bar{A}$ satisfying (a) i.e.,  $\eta(x, \gamma_{A}(x)) = 1/n$ for all $x \in X(A)$, the bipartite state $\eta^{t}$ defined by $\eta^{t}(x,\gamma(y)) = \eta(y,\gamma(x))$ is also perfectly uniformly correlating, whence, so is the symmetic state $(\eta + \eta^{t})/2$. 
We therefore can, and do, assume in what follows that {\em the chosen correlating state $\eta_A$ is always symmetric.} %\footnote{In fact, 
If $A$ is sharp, it is easy to show that $\eta$ is uniquely determined by condition (a) of Definition 1: since $\eta(x,\bar{y}) = 0$ for outcomes $y \not = x$ belonging to a common test, and $\eta(y,\bar{y}) = 1/n$, we have $\eta_{1|\bar{y}}(y) = 1$, i.e,. 
$\eta_{1|\bar{y}} = \delta_{y}$ where $\delta_y$ is the unique state in which $y$ has probability $1$. Thus, $\eta(x,\bar{y}) = \delta_{y}(x)$.  
In this case, therefore, $\eta = \eta^{t}$, i.e., $\eta$ is automatically symmetric.\footnote{It 
follows that, where $\delta_x$ and $\delta_y$ are the unique states making $x$ and $y$ certain, 
we have $\delta_{y}(x) = \delta_{x}(y)$ for all $x, y \in X(A)$. Indeed, this last condition could substitute for condition (a) in the definition of  $\eta$, as it implies that $\eta(x,\bar{y}) := \delta_{x}(y)$ is a valid non-signaling probabilty weight on $A$ and $\bar{A}$.}
% {\blue Ref 2 finds this footnote opaque. Have we defined $\delta_x$ yet??}}
%}

If $A(\H)$ is the quantum probabilistic model associated with a finite-dimensional Hilbert space, then the EPR state $\Psi$ turns $\bar{A}(\H) := A(\bar{\H})$ into a 
conjugate in this sense, with $\eta_{A}(x,\bar{y}) = |\langle \Psi, x \otimes \bar{y} \rangle|^2$.
%\langle \Psi x \otimes \bar{y}, x  \otimes \bar{y} \rangle$.  
In fact, as pointed out in the introduction, $A(\bar{\H})$ is a conjugate for $A(\H)$, since every density operator is the marginal of a state $\Psi_W$ correlating 
an eigenbasis for $W$ with its conjugate. 
All of this works equally well for real quantum systems, taking $\bar{\H} = \H$. With a little care, it can be shown to work for quaternionic systems as well.\\

\noindent{\em Remark:} 
One might wonder whether one can use the isomorphism $\gamma_{A}$ to identify $A$ with its conjugate. Certainly one can define a bipartite state $\eta_{A}'(x,y) = \eta_{A}(x,\gamma_{A}(y))$. However, whether this corresponds to a legitimate state of any physically 
reasonable composite $AA$ of $A$ with itself, depends on the particular probabilistic theory 
at hand. For example, if $A = A(\H)$ is the quantum model associated with a complex Hilbert space $\H$, then $(a,b) \mapsto \Tr(ab)$ corresponds to 
{\em no} state on $A(\H \otimes \H)$. On the other hand, any choice of an anti-unitary operator $J$ acting on $\H$ yields a unitary isomorphism 
$\bar{J} : \H \rightarrow \bar{\H}$, {\blue given by $\bar{J}(x) = \bar{J(x)}$ for all $x \in \H$}. Defining $\eta'(x,y) := |\langle (\1 \otimes \bar{J})^{-1}\Psi, x \otimes y \rangle|^2 = |\langle \Psi, x \otimes \bar{Jy} \rangle|^2$ gives a state on $A(\H \otimes \H)$, correlating along the anti-unitary isomorphism $\gamma'(x) := J^{-1}x$.  Thus, whether we choose to treat $A(\H)$ as its own conjugate, or as distinct from its conjugate, is to an extent a matter of convention. \\

\tempout{\footnote{ One might wonder whether one can use the isomorphism $\gamma_{A}$ to simply identify $A$ with its conjugate. Certainly we can use $\gamma_{A}$ to pull the 
bipartite state $\eta_{A}$ on $A \bar{A}$ back to a positive bilinear form on $\E(A) \times \E(A)$, namely $\eta_{A}'(a,b) = \eta_{A}(a,\gamma_{A}(a))$. However, whether this corresponds to a legitimate bipartite {\em state} on a legitimate composite $AA$ of $A$ with itself, depends on the particular probabilistic theory 
at hand. For example, if $A = A(\H)$ is the quantum model associated with a Hilbert space $\H$, and $\bar{A}$ is the model associated with $\bar{\H}$ in the usual way, then the state $\eta_A$ 
pulls back along the isomorphism $a \mapsto \bar{a}$ to a bilinear form on $\L_{\sa}(\H) \times \L_{\sa}(\H)$ --- but one associated with the non-completely positive partial-transpose operation, which corresponds to no state on $A(\H \otimes \H)$. Thus, the notion of a conjugate system is best understood as 
applying to an entire probabilistic theory, rather than to a single probabilistic model.}}

\noindent{\bf Conjugates and Self-Duality} We are now ready to prove Theorem 1, which, for convenience, I restate. Recall that a model $A$ is {\em sharp} iff, for every outcome $x \in X(A)$, there is a unique state 
$\delta_x \in \Omega(A)$ with $\delta_x(x) = 1$. Both classical and quantum models are sharp. \\

\noindent{\bf Theorem 1 (bis)} {\em Let $A$ be %uniform, 
sharp and have a conjugate $\bar{A}$. Then 
$\langle a, b \rangle := \eta_{A}(a,\gamma_A(b))$ is a self-dualizing inner product on $\E(A)$, and 
induces an order-isomorphism $\E(A) \rightarrow \V(A)$ given by $a \mapsto \eta_{A}(a, \ \cdot \ )$ for $a \in \E(A)$. 
\footnote{Here, I am identifying $\V(A)$, as a vector space, with $\E(A)^{\ast}$.}\\
}
 
The proof is not difficult. It will be convenient to break it up into a sequence of even easier lemmas. In the interest of readability, 
below I conflate $x \in X(A)$ with the corresponding effect $\hat{x} \in \E(A)$, 
and write $\bar{x}$ for $\gamma_{A}(x)$. 
Until further notice, the hypotheses of Theorem 1 are in force. 
 The first step is to obtain a kind of weak ``spectral" decomposition for states in $\Omega(A)$ in terms of the states $\delta_x$. %\\

\begin{lemma} For every $\alpha \in \V(A)_+$, there exists a test $E$ 
such that 
\begin{equation} \alpha \ = \ \sum_{x \in E} \alpha(x) \delta_{x}\end{equation}
\end{lemma}

\noindent{\em Proof:} We can assume that $\alpha$ is a normalized state, i.e, that $\alpha \in \Omega(A)$. Since $\bar{A}$ is a conjugate for $A$, $\alpha = \omega_1$ where $\omega$ correlates some test $E \in \M(A)$ with $\bar{E} \in \M(\bar{A})$ along the bijection $x \mapsto \bar{x}$. By the law of total probability (1) for non-signaling states, 
$\alpha = \sum_{\bar{x} \in \bar{E}} \omega_2(\bar{x})\omega_{1|\bar{x}}$. 
 Since $\omega$ is correlating we have $\omega_{1|\bar{x}}(x) = 1$. Thus, by sharpness, $\omega_{1|\bar{x}} = \delta_{x}$. Hence, 
$\alpha = \sum_{x \in E} \omega_{2}(\bar{x}) \delta_{x}$. It follows that $\omega_2(\bar{x}) = \alpha(x)$ 
for every $x \in E$, giving us (\theequation). 
$\Box$\\

We will refer to the decomposition in equation (4) as a {\em spectral decomposition} for $\alpha$. 
%and say that a model in which such a decomposition is available for all states is {\em spectral}. \\

\tempout{
\noindent{\em Remark:} This is the only point in the derivation in which we need to make use of condition (a) in the definition of a conjugate. Thus, if we are willing to take the existence of ``spectral" decompositions of the form (\theequation) for granted, as in \cite{BMU}, 
it suffices to assume only the existence of a {\em weak} conjugate, i.e., the existence of a universally and uniformly correlating bipartite state $\eta$ on $A$ and $\bar{A}$.  
}

\begin{lemma}  $\eta_{A}$ is an isomorphism state. 
\end{lemma}
 
\noindent{\em Proof:} We need to show that $\hat{\eta_{A}} : \E(A) \rightarrow \V(A)$ is 
an order-isomorphism. Since $\E(A)$ and $\V(A)$ have the same dimension,  %a linear isomorphism with a positive inverse. 
it is enough to show that $\hat{\eta_{A}}$ maps the positive cone of $\E(A)$ {\em onto} that of $\V(\bar{A})$.  Since $x \mapsto \bar{x}$ is an isomorphism between 
$A$ and $\bar{A}$, we can apply Lemma 1 to $\bar{A}$: if $\alpha \in \V_{+}(\bar{A})$, we have $\alpha = \sum_{x \in E} \alpha(\bar{x}) \delta_{\bar{x}}$. 
Since $\eta_{A}(x,\bar{x}) = 1/n$, we have $\hat{\eta_{A}}(x) = \frac{1}{n} \delta_{\bar{x}}$ for every $x \in X(A)$. 
Hence, $\hat{\eta_{A}}\left ( \sum_{x \in E} n \alpha({x}) x \right )  = \alpha$. $\Box$ 

\begin{lemma} Every $a \in \E(A)$ has a representation $a = \sum_{x \in E} t_x x$ for some test $E \in \M(A)$ and some coefficients $t_x$. 
\end{lemma}

\noindent{\em Proof:} If $a \in \E(A)_+$, then by Lemma 1, 
$\hat{\eta_{A}}(a)  
 = \sum_{x \in E} t_x \delta_{\bar{x}}$ 
 for some $E \in \M(A)$ and coefficients $t_x \geq 0$. By Lemma 2, $\hat{\eta}_{A}$ is an order-isomorphism. Applying $\hat{\eta}_{A}^{-1}$ to the expansion above % (\theequation) 
gives $a = \sum_{x \in E} t_x x$. %the desired result. %expansion of $a$.
For an arbitrary $a \in \E(A)$, we have $a = a_1 - a_2$ with $a_1, a_2 \in \E(A)_+$. Choose $N  \geq 0$ with $a_2 \leq Nu$. Thus, $b := a + Nu_A = a_1 + (Nu_A - a_2) \geq 0$.  So $b := \sum_{x \in E} t_x x$ for some $E \in \M(A)$ and thus 
\[\ \ \ a = b - Nu_A = \sum_{x \in E} t_x x - N(\sum_{x \in E} x) = \sum_{x \in E} (t_x - N)x. \  \Box\]

%\noindent{\em Remark:} Lemma 3 is of independent interest, since it allows us to regard every element of $\E(A)$ as 
%``physically meaningful". 

\begin{lemma} The function $\langle a, b \rangle := \eta_{A}(a, \gamma_{A}(b))$ is an inner product on 
$\E(A)$.
\end{lemma}

\noindent{\em Proof:} $\eta_{A}$ is bilinear and, by assumption, symmetric. 
%It's also bilinear, because $\eta_{A}$ is non-signaling and $\gamma_A$ is linear. 
We need to show that $\langle ~,~ \rangle$ is positive definite. Let $a \in \E(A)$. From Lemma 3, we have $a = \sum_{x \in E} t_x x$ for 
some test $E$ and coeffcients $t_x$.  Now 
\[
\langle a, a \rangle 
= \left \langle \sum_{x \in E} t_x x, \sum_{y \in E} t_y y \right \rangle 
 = \sum_{x, y \in E \times E} t_x t_y \langle x, y \rangle 
 =  \sum_{x, y \in E \times E} t_x t_y \eta_{A}(x, \bar{y}) 
 =  \frac{1}{n} \sum_{x \in E} {t_{x}}^2 \geq 0.
\]
\tempout{
\begin{eqnarray*}
\langle a, a \rangle 
= \left \langle \sum_{x \in E} t_x x, \sum_{y \in E} t_y y \right \rangle 
& = & \sum_{x, y \in E \times E} t_x t_y \langle x, y \rangle \\
& = & \sum_{x, y \in E \times E} t_x t_y \eta_{A}(x, \bar{y}) \\
& = & \frac{1}{n} \sum_{x \in E} {t_{x}}^2 \geq 0.
\end{eqnarray*}
}
This is zero only when all coefficients $t_x$ are zero, i.e., only for $a = 0$.  $\Box$\\

\noindent{\em Proof of Theorem 1, concluded:} 
Lemma 2 tells us that $\E(A) \simeq \V(A)$,  %Hence, %to prove Theorem 1, 
so it remains only to show that the inner product $\langle \ , \ \rangle$ is self-dualizing. Clearly $\langle a, b \rangle = \eta(a, \bar{b}) \geq 0$ for all $a, b \in \E(A)_+$. 
Suppose $a \in \E(A)$ is such that $\langle a, b \rangle \geq 0$ for all 
$b \in \E(A)_+$. Then $\langle a, y \rangle \geq 0$ for all $y \in X$. By Lemma 3, $a = \sum_{x \in E} t_x x$ for some test $E$. Thus, for all $y \in E$ we have 
$\langle a,y \rangle = t_y \geq 0$, whence, $a \in \E(A)_+$. $\Box$ \\

\tempout{
Suppose now that every non-singular state of $A$ can be prepared, up to normalization, from the maximally mixed state $\rho(x) \equiv 1/n$ by some reversible process. This guarantees that $\V(A)$, and hence, $\E(A)$, is  homogeneous. 
\tempout{Indeed, as discussed above, a reversible process is an order-automorphism $\phi : \V(A) \rightarrow \V(A)$; 
to say that this prepares $\alpha$, up to normalization, from $\rho$, is simply to say that $\alpha = t\phi(\rho)$ for some appropriate constant $t > 0$ (namely, $t = u_{A}(\phi(\rho))^{-1}$). 
Since $t\phi$  is again an order-automorphism, it follows that the group of order-automorphisms of $\V(A)$ act transitively on the interior of $\V(A)_+$, i.e., $\V(A)$ is homogeneous. }
By the Koecher-Vinberg Theorem, $\E(A)$ carries a unique euclidean Jordan structure making 
$\E(A)_+$ the cone of squares and $u_A$, the Jordan unit. \\}

\noindent{\em Remarks} There are several directions in which we can usefully modify the assumptions of Theorem 1.

(1) In the proof of Theorem 1, the only point at which we needed to assume that $\bar{A}$ satisfies condition (b) in the definition of a conjugate was in order to obtain the spectral decomposition --- equation (4) --- 
of Lemma 1. Thus, {\blue if we are willing simply to {\em assume} such decompositions are available, as in \cite{BMU}, 
then a weak conjugate suffices. Alternatively, {\em any} postulate or postulates leading to such decompositions can replace 
condition (b). For instance, certain versions of the symmetry and  ``subspace" axioms used in 
\cite{Hardy, Dakic-Brukner, Masanes-Mueller} imply a spectral decomposition. This is spelled out in Appendix B.  Another approach to obtaining such a decomposition can be 
found in a recent paper of G. Chiribella and C. M. Scandolo \cite{Chiribella-Scandolo}. }

 (2)  In fact,  it is even enough if (4) holds for states in the {\em interior} of $\Omega(A)$. 
From this we have, as in the proof of Lemma 2, that the interior, $\V^{\circ}_{+}$, of the cone $\V(A)_+$ is contained in $\hat{\eta}_A(\E_+)$, from which it follows that $\hat{\eta}$ is a linear isomorphism, and hence (as 
the vector spaces involved are finite-dimensional) an homeomorphism. Thus, $\hat{\eta}_{A}(\E_+)$ is closed, 
and so, contains the closure of $\V_{+}^{\circ}$, i.e., $\V_{+}$. In other words, $\hat{\eta}$ is an 
order-isomorphism. The proofs of Lemmas 3 and 4, and the rest of the proof of Theorem 1, then proceed just as 
before.

(3) The definition of a conjugate for a probabilistic model $A$ requires the existence of the uniformly, universally correlating state $\eta_A$, {\em and} that arbitrary states of $A$ arise as marginals of bipartite states on $A$ and $\bar{A}$ correlating {\em some} test $E\in \M(A)$ with its conjugate twin. One might wonder whether there is some reasonably simple postulate that will imply both of these conditions. Suppose that $G$ is a group acting transitively on the outcome-space $X(A)$ of the model $A$, and leaving the state-space $\Omega(A)$ invariant. If $G$ is compact, there will exist an invariant state, $\rho$, obtained by group averaging; by the transitivity of $G$ on outcomes, this state must be constant. It follows that all tests 
have the same finite size size, say $n$,  and that $\rho$ is the maximally mixed state $\rho(x) \equiv 1/n$. 
That is, the model is uniform. 
Now let $\gamma_A : x \mapsto \bar{x}$ be an isomorphism between 
$A$ and a model $\bar{A}$. 
Suppose that every state $\alpha \in \Omega(A)$ is the marginal of a correlating state $\omega \in \Omega(A\bar{A})$ such 
that $\omega(gx, \bar{gy}) = \omega(x,y)$ for all $g \in G$ with $\alpha \circ g = \alpha$. 
It is easily checked that this is satisfied by finite-dimensional quantum models. Applied to the maximally mixed state $\rho$, this produces a perfectly, uniformly correlating state $\eta_A$. Thus, $\bar{A}$ is a conjugate in the sense of Definition 3.\\

\tempout{(4) It is certainly reasonable to add to the structure of a probabilistic model $A$ a distinguished group $G(A)$ of ``physical symmetries" acting on $X(A)$ so as to map tests to tests, also requiring that if $\alpha \in \Omega(A)$, 
$\alpha \circ g \in \Omega(A)$ as well for each $g \in G(A)$. Indeed, such a privileged symmetry group is an important part of the framework assumed in most other recent reconstructions of QM.  Suppose we are given 
such a group $G(A)$, where $A$ has a weak conjugate $\bar{A}$. Then $G(A)$ acts on $\bar{A}$ in an obvious way. For any fixed $g \in G(A)$ the function $(x,y) \mapsto \eta_{A}(gx, \bar{gy})$ defines a weak conjugate on $A$. 
If $A$ is sharp, the correlator is unique. Hence, $\eta_{A}(gx, \bar{gy}) = \eta_{A}(x,\bar{y})$ for all $x, y \in X(A)$. Equivalently, $\eta_{A}(gx, \bar{y}) = \eta_{A}(x, \bar{g^{-1} y})$. If we understand a {\em dynamics} 
for $A$ to be a one-parameter subgroup of $G(A)$, i.e., a homomorphism $(\R,+) \rightarrow G(A)$ given by $t \mapsto g_{t}$, where we understand $g_{t}$ to be the translation of the system forward in time by $t$ units, 
it follows that $\bar{A}$ is a {\em time reversed} version of $A$, for {\em all possible} dynamics.  [NO: put in CONCLUSION!]} 

\section{Filters}

We have just seen that if $A$ is sharp and has a conjugate, then $\E(A)$ is self-dual, and isomorphic to $\V(A)$. 
Suppose now that every non-singular state of $A$ can be prepared, up to normalization, from the maximally mixed state $\rho(x) \equiv 1/n$ by some reversible process. This guarantees that $\V(A)$, and hence, $\E(A)$, is  homogeneous, 
so that, by the Koecher-Vinberg Theorem, $\E(A)$ carries a euclidean Jordan structure making 
$\E(A)_+$ the cone of squares.
\tempout{Indeed, as discussed above, a reversible process is an order-automorphism $\phi : \V(A) \rightarrow \V(A)$; 
to say that this prepares $\alpha$, up to normalization, from $\rho$, is simply to say that $\alpha = t\phi(\rho)$ ]for some appropriate constant $t > 0$ (namely, $t = u_{A}(\phi(\rho))^{-1}$). 
Since $t\phi$  is again an order-automorphism, it follows that the group of order-automorphisms of $\V(A)$ act transitively on the interior of $\V(A)_+$, i.e., $\V(A)$ is homogeneous. }
 % and $u_A$, the Jordan unit. %\\

In fact, we can say something more interesting.  
In many kinds of laboratory experiments, the distinct outcomes of an experiment correspond to physical detectors, 
the efficiency of which can independently be attenuated, if desired, by the experimenter. This can always be done through post-processing, using a classical filter. In QM, it can also be accomplished 
by subjecting the system to a suitable process {\em prior} to measurement. 
To see this, let $A$ be a finite-dimensional quantum system, with corresponding Hilbert space $\H$, and identify 
$\E(A)$ with $\L_{\sa}(\H_A)$. If $E$ is an orthonormal basis representing a basic measurement on this system, define a positive operator $V : \H \rightarrow \H$ by setting $Vx = t_{x}^{1/2} x$ for every $x \in E$, where $0 \leq t_x \leq 1$. This gives us a completely positive linear mapping $\Phi : \E(A) \rightarrow \E(A)$, namely $\Phi(a) = VaV$. 
If $t_x > 0$ for every $x \in E$, $\Phi$ has a completely positive inverse $\Phi^{-1}(a) = V^{-1}a V^{-1}$. For each $x \in E$, the corresponding effect $\hat{x} \in \E(A) \simeq \L_{\sa}(\H)$ is the rank-one projection operator $p_x$. It is easy to check that $V p_x V = t_{x} p_x$, i.e., that $\Phi(\hat{x}) = t_x \hat{x}$ for every $x \in E$. %\\

%\newpage
\begin{definition} A {\em filter} for a test $E$ of a probabilistic model $A$ is a positive linear mapping $\Phi : \V(A) \rightarrow \V(A)$ such that, for every outcome $x \in E$, there exists a coefficient 
$0 \leq t_x \leq 1$ with % for some coefficients $0 \leq t_x \leq 1$, 
\[\Phi(\alpha)(x) = t_x \alpha(x)\] 
for all all states $\alpha \in \Omega(A)$. Equivalently, $\Phi^{\ast}(x) = t_x x$ for every $x \in E$. %\\
\end{definition} 
%In the ``beams and detectors" model, the special case in which $t_x$ is either $0$ or $1$ for all $x \in E$ corresponds to %the case in which the filter either allows or blocks passage of a system through each detector.
As noted above, in QM, not only do filters with arbitrary coefficients exist for every test, but they can be implemented $p$-reversibly, so long as the coefficients $t_x$ are all non-zero. I will say that a general probabilistic model with this feature {\em has arbitrary reversible filters}. \\

\noindent{\bf Corollary 1 (bis)} {\em Suppose that $A$ is sharp and has a conjugate $\bar{A}$. If $A$ has arbitrary reversible filters, then $\E(A)$ is homogeneous and self-dual.} \\

\noindent{\em Proof:} $A$ is self-dual by Theorem 1. Let $\alpha$ be a normalized state in the interior of $\V(A)_+$. By Lemma 1, $\alpha$ has a spectral 
decomposition $\alpha = \sum_{x \in E} \alpha(x) \delta_{x}$. Let $\Phi$ be a filter for $E$ with coefficients 
$\alpha(x)$.  Since $\alpha$ is non-singular, $\alpha(x) > 0$ for 
all $x \in E$, so $\Phi$ can be chosen to be reversible. Now expand the maximally mixed state $\rho$, with 
$\rho(x) \equiv 1/n$, as $\rho = \sum_{x \in E} \frac{1}{n} \delta_x$. Then 
$\Phi(\rho) = \frac{1}{n} \sum_{x \in E} \alpha(x) \delta_x = \frac{1}{n} \alpha$. 
Thus, any non-singular state can be prepared, up to normalization, by a reversible filter, and it follows that $\V(A)$ is homogeneous. In view of Theorem 1, $\E(A)$ is self-dual, and 
$\E(A) \simeq \V(A)$, whence, also homogeneous. $\Box$ \\

\noindent{\bf State preparation by reversible filters}  Suppose now that $A$ has only 
a {\em weak} conjugate $\bar{A}$, and that $\Phi$ is a filter for a test $E \in \M(A)$. By applying $\Phi$ to one of the two systems $A$ and $\bar{A}$, 
% wing of the composite system $A\bar{A}$,
 we can convert the correlator $\eta_A$ into a new sub-normalized bipartite state  
$\omega$, given by $\omega(x,y) = \eta_A(\Phi^{\ast} x, y)$ for all $x \in X(A), y \in X(B)$. 
Noticing that $\Phi^{\ast}(x) = t_x x$ for every 
$x \in E$, we see  that $\omega$   
%$\omega = \eta_{A} \circ (\phi \otimes \bar{\1})$ 
correlates $E$ with $\bar{E}$: if $x, y \in E$ with $x \not = y$, we have 
\[\omega(x,\bar{y}) = \eta_{A}(t_x x, \bar{y}) = t_{x}\eta_{A}(x,\bar{y}) = 0.\] 
In other words, $\omega$ is correlating. It follows that the normalized bipartite 
state 
\[\tilde{\omega} := \frac{1}{\omega(u_A, u_{\bar{A}})} \omega\]
is likewise correlating. Since $\omega_1 = \Phi(\rho)$, it follows that {\em any state preparable from $\rho$ by a filter} --- that is, any state of the form $\alpha = \tilde{\Phi(\rho)}$, where $\Phi$ is a filter and 
${\blue \tilde{\alpha} := \tfrac{1}{u_{A}(\alpha)}\alpha}$ ---  is the marginal of a correlating 
state, and hence enjoys a spectral decomposition as in Equation {\blue (4)}.
Thus, if every state is so preparable,  the weak conjugate $\bar{A}$ 
is actually a conjugate. So, in the presence of sharpness, we can replace the assumption that the conjugate is strong, by the requirement that {\em every} state be preparable by a filter. 
In fact, by strengthening this preparability assumption, it is even possible to omit the hypothesis that $A$ is sharp. 
% we can do a bit better. 
%{\bf [This could be clearer re normalization.]}

The isomorphism $\gamma_A : A \simeq \bar{A}$ extends to an order-automorphism $\V(A) \simeq \V(\bar{A})$, given by $\alpha \mapsto \bar{\alpha}$, with $\bar{\alpha}(\bar{x}) = \alpha(x)$ for all $x \in X(A)$. Hence, a positive linear mapping $\Phi : \V(A) \rightarrow \V(A)$ has a counterpart $\bar{\Phi} : \V(\bar{A}) \rightarrow \V(\bar{A})$, given by $\bar{\Phi}(\bar{\alpha}) = \bar{\Phi(\alpha)}$.  
Let us say that $\Phi$ is {\em symmetric} with respect to $\eta_A$ iff $\eta_A(\Phi^{\ast}(x),\bar{y}) = \eta_A(x,\bar{\Phi^{\ast}}(y))$ for all $x, y \in X(A)$, i.e., iff $\eta_A \circ (\Phi^{\ast} \otimes \id_{\bar{A}}) = \eta_A \circ (\id_{A} \otimes \bar{\Phi}^{\ast})$. \\

%\newpage
\noindent{\bf Lemma 5} {\em Let $A$ have a weak conjugate $\bar{A}$. Suppose that every state of $A$ is preparable by a symmetric filter. Then  $\langle a, b \rangle := \eta_{A}(a, \gamma_{A}(b))$ is a self-dualizing inner product on $\E(A)$.}\\ %with respect to which $\E(A)$ is self-dual.} 

\noindent{\em Proof:} Let $\alpha = \Phi(\rho)$, where $\Phi$ is a symmetric filter for some test $E$. Consider the bipartite state 
\[\omega  := \eta_A \circ (\Phi^{\ast} \otimes \id_{\bar{A}})  = \eta_{A} \circ (\id_{A} \otimes \bar{\Phi}^{\ast}).\]
For each outcome $x \in X(A)$, let 
%\[\delta_x := (\eta_{A})_{1|\bar{x}}.\]
$\delta_{x}$ denote the conditional state $(\eta_{A})_{1|\bar{x}}$. 
Then for all $x \in E$, and all 
outcomes $y \in X$, we have  
\begin{eqnarray*}
\omega_{1|\bar{x}}(y)  =  \frac{\eta_{A}(\Phi^{\ast} (y), \bar{x})}{\eta_{A}(\Phi^{\ast}(u_{A}),\bar{x})}
 & = & \frac{\eta_{A}(y,\bar{\Phi^{\ast}}(\bar{x}))}{\eta_{A}(u_{A},\bar{\Phi^{\ast}}(\bar{x}))} \\
 & = & \frac{\eta_{A}(y,t_x \bar{x})}{\eta_A (u_A ,t_{x}\bar{x})} \\
 & = & \frac{\eta_{A}(y,\bar{x})}{\eta_{A}(u_A ,\bar{x})} = (\eta_{A})_{1|\bar{x}}(y) = \delta_{x}(y).
\end{eqnarray*}
It follows that $\omega_{1|\bar{x}} = \delta_{x}$. It is easy to check that %%DO SO??
$\omega_1 =  \Phi((\eta_{A})_1) = \Phi(\rho) = \alpha$;  
also, by the law of total probability (3), $\omega_1 =  \sum_{x \in E} \omega_{2}(\bar{x}) \omega_{1|\bar{x}} = \sum_{x \in E} t_x \delta_x$, 
where $t_x = \omega_{2}(\bar{x})$. 
Thus, every state in $\Omega(A)$ is a 
convex combination of the states $\delta_x$, and the cone generated by these states coincides with $\V(A)_+$. 
It follows that $\hat{\eta}$ maps $\E(A)_+$ onto $\V(A)_+$, as in the proof of Lemma 2.  
The proof that $\langle a, b \rangle := \eta(a, \bar{b})$ defines an inner product on $\E(A)$ now proceeds 
as in the proof of Lemmas 3 and 4. $\Box$\\

In fact, we can do a bit better: \\

\noindent{\bf Corollary 2 (bis)} {\em Let $A$ have a weak conjugate, and suppose that every {\em interior} state is preparable by a {\em reversible} symmetric filter. Then $A$ is homogeneous and self-dual.}\\

\noindent{\em Proof:} The preparability assumption clearly makes $\V(A)$ homogeneous. The proof of Lemma 5 
shows that all states in the interior of $\Omega$ can be decomposed as in equation (4) with respect to the states $\delta_x = \eta_{1|\bar{x}}$. 
As noted in Remark (2) following the proof of Theorem 1, this is enough to secure the self-duality of $\E(A)$, and its isomorphism with $\V(A)$. $\Box$ 
\\

It follows from the KV theorem that, for any model $A$ satisfying the hypotheses of either Corollary 1 or Corollary 2, 
$\E(A)$ carries a Jordan product compatible with the inner product arising from $\eta_{A}$, i.e, $\E(A)$ is a 
euclidean Jordan algebra.  In fact, one can prove more: the unit effect $u$ coincides with the Jordan unit, and 
$\M(A)$ is precisely the set of Jordan frames. In other words, $\E(A)$ is a Jordan moel.  The proof is given in Appendix A, where it is also shown that any Jordan model satisfies the hypotheses of both corollaries. Thus, these two sets of hypotheses are equivalent, and exactly characterize the class of Jordan models. 
To summarize:\\

\noindent{\bf Theorem 2} {\em For a finite-dimensional, uniform probabilistic model $A$, the following statements are equivalent: 
\begin{mlist} 
\item[(a)] $A$ is sharp, has a conjugate, and has arbitrary reversible filters
\item[(b)] $A$ has a weak conjugate, and all non-singular states can be prepared by reversible symmetric filters 
\item[(c)] $A$ is a Jordan model.\\
\end{mlist}}

It should be stressed that all of the assumptions going into (a) and (b) are what \cite{BMU} calls {\em single-system} postulates, at least to the extent that the existence of a conjugate (or weak conjugate) is a property of a single 
system. In any event, these assumptions, whether seen as pertaining to a single system $A$ or to the pair 
$(A,\bar{A})$, are quite different in flavor from local tomography or the subspace axiom, which place constraints 
on an entire {\em theory's worth} of probabilistic models. 

\section{Conclusion} 

We've seen that either of two related packages of assumptions --- given in (a) and (b) of Theorem 2 --- 
 lead in a very simple way the homogeneity and self-duality of the space $\E(A)$  associated with a probabilistic model $A$, and hence, by the Koecher-Vinberg Theorem, to $A$'s having a euclidean Jordan structure. While this is not the only route one can take to deriving this structure (see, e.g, \cite{MU} and \cite{SSD} for approaches stressing symmetry principles), it does seem especially straightforward.

As discussed in the introduction, several other recent papers (e.g, \cite{Hardy, Rau, Dakic-Brukner, Masanes-Mueller, CDP}) have derived standard finite-dimensional quantum mechanics, over $\mathbb C$, from operational or information-theoretic axioms. Besides the fact that the mathematical development here is quicker and easier, 
the axiomatic basis is considerably different, and arguably leaner, making no appeal to the structure of subsystems, 
or to the isomorphism of systems with the same information-carrying capacity, or to local tomography. The last 
two points are particularly important: by avoiding local tomography, we allow for real and quaternionic 
quantum systems; by not insisting that physical systems having the same information capacity be isomorphic, 
we allow for quantum theory with superselection rules, and for physical theories in which real and quaternionic systems can coexist. Of course, the door has been opened a bit wider than this: our postulates are also compatible with spin 
factors and with the exceptional Jordan algebra.\footnote{It can be shown \cite{BGraW} that the exceptional 
Jordan algebra  can be ruled 
out on the grounds that one can form no satisfactory composite of two euclidean Jordan algebras if either has an 
exceptional direct summand. Whether the spin factors can also be discarded, or whether they have some physical role to play, remains an open question.}

Of the reconstructions cited above, the one having the strongest affinity with the approach of this paper is 
that of \cite{CDP}, 
the key postulate of which is that every state dilates to --- that is, arise as the marginal of --- a  {\em pure} state on a larger, composite system, unique up to symmetries of the ancillary system. Condition (a) in the 
definition of a conjugate, requiring that every state dilate to a correlating state, has a somewhat similar character, albeit with the emphasis on the dilated state's correlational properties, rather than its purity. 
To make the connection more explicit, suppose we require that every {\em non-singular} state $\alpha$ on $A$ dilate to a correlating {\em isomorphism} state 
$\omega$ (which is the case, in the presence of our other assumptions). 
If $\mu$ is another isomorphism %correlating 
state with the same marginal state $\alpha$, then $\phi := \hat{\mu} \circ \hat{\omega}^{-1}$ is a 
reversible transformation on $\V(\bar{A})$ with $\hat{\mu} = \phi \circ \hat{\omega}$, i.e., $\mu(a,b) = \omega(a,\phi(b))$ for all $a, b \in \E(A)$. 
Now, as shown in \cite{BGW}, if $\V(A)$ is irreducible as an 
ordered vector space, isomorphism states are pure. Thus, in the irreducible case, we have a 
version of the purification postulate for non-singular states.  
In view of these connections, it seems plausible that the approach taken here might be adapted to considerably simplify the mathematical development in \cite{CDP}. \footnote{\blue Going in the other direction, in \cite{Chiribella-Scandolo}, the authors derive a version of part (a) of our definition of a conjugate, that is, the existence of a dilation perfectly correlating two tests, from  axioms similar to those of \cite{CDP}.  More recently, in the conext of a compact closed category 
of processes, the authors of \cite{Selby-Scandolo-Coecke} introduce a 
stronger, ``symmetric" version of the purification postulate, and show that when combined with suitable versions 
of sharpness and the existence of a ``classical interface", this implies that all states 
can be prepared from the maximally mixed state by a reversible process, allowing them to prove that their analogue of the cone $\V(A)_+$ is homogeneous and self-dual.}
 
%{\blue [AND ELSEWHERE. And perhaps this is the place for more comment 
%on symmetric purification?]}

{\blue An assumption that is common to nearly all of the cited earlier reconstructions is some version of 
Hardy's {\em subspace postulate}, which requires (roughly speaking) that the result of constraining a 
physical system to the set of states making a particular measurement-outcome impossible, also count 
as a physical system.  This very powerful assumption, while not needed in the development above, can 
readily be adapted to the framework of this paper, and can to a large extent replace our assumptions 
above about the existence of reversible filters. The details can be found in Appendix B.} 

%\tempout
{\blue 
%\noindent{\bf Jordan algebraic probabilistic theories} 
It is worth remarking that the subspace axiom applies, not to an indidual probabilistic model but to a {\em class} of probabilistic models, 
that is, to an entire {\em probabilistic theory}. (In the language of \cite{BMU}, it is not a {\em single-system} postulate.)   
%However, there is nothing 
%at all {\em wrong} with such ``theory-wide" axioms. In fact, 
As a rule, 
one wants to think of a physical theory, not as a loosely structured class, but as a {\em category} of systems, 
with morphisms corresponding to processes. To allow for composite systems, it is natural to take this to be a 
{\em symmetric monoidal category} \cite{Abramsky-Coecke}. This brings us to the interesting 
question of whether one can construct symmetric monoidal categories of  probabilistic models, 
in which (say) the hypotheses of Corollary 1 are satisfied by all systems. This 
is indeed possible for {\em special} Jordan algebras (those not having the exceptional Jordan algebra as a direct 
summand). Restricting attention to Jordan models corresponding to direct sums of real, complex and quaternionic matrix algebras, one can even arrange for this category to be compact closed 
\cite{BGW}. This implies that many standard quantum-information theoretic protocols, notably conclusive 
teleportation and entanglement-swapping, are still available in this non-locally tomographic setting.
\footnote{More 
generally, the existence of conjugate systems is very suggestive of the ``caps" that define part of a compact 
structure on a symmetric monoidal category (an observation reflected in my choice of the notation $\eta$). 
For a further development of this connection, see \cite{Shortcut}.  See also \cite{Selby-Scandolo-Coecke} for a 
reconstruction of finite-dimensional quantum theory from a ``symmetric purification" postulate on the 
structure of a suitable dagger-monoidal category allowing ``classical control".)}\\}

\tempout{As a final comment, one would like a clearer view of the physical meaning of both strong and weak conjugates. The place to look for guidance, of course, is ordinary finite-dimensional QM. If $\H$ is a complex Hilbert space, corresponding to a 
quantum mechanical system, what, in general, is the physical significance of the conjugate space $\bar{\H}$? Unfortunately, there seems to be little consensus about this. Above, I've 
suggested that it represents a kind of ``storage medium" where records of measurement outcomes are recorded as states. This is certainly not the only possible reading.  
One can also regard $\bar{\H}$ as representing the same system, but with ``input" and ``output" --- preparation of states and registration of outcomes --- reversed. Similarly (though this is not quite the same thing), one can regard $\bar{\H}$ as representing a time-reversed version of the system corresponding to $\bar{\H}$. 
It is not clear to me that any of these interpretations gives the full story, but in any case, the results obtained here strongly suggest that it would be profitable to find an simple, general physical interpretation of conjugate {\em quantum} systems that {\em accounts} for the existence of the perfectly, uniformly correlating EPR state. \\}

\tempout{
Even so, one wants to say more about the physical (or operational, or probabilistic, or information-theoretic) interpretation of the conjugate system. 
In quantum theory, one can regard the conjugate Hilbert space $\bar{\H}$ as representing a time-reversed version of the system represented by $\H$, and a similar interpretation is available for conjugates in the general setting of 
this paper. A detailed 
development of this idea and its consequences will appear elsewhere.\\ % be the subject of a later paper.\\
}

\noindent{\large \sf Acknowledgements} 
%\thanks{
I wish to thank Giulio Chiribella, Chris Heunen, Matt Leifer and Markus M\"{u}ller for helpful comments on earlier drafts
of this paper. This work was  supported in part by a grant (FQXi-RFP3-1348) from the FQXi foundation.
%}

%\bibliography{CFQMrev3}
%\bibliographystyle{plainnat}

\appendix
\setcounter{lemma}{0}
\renewcommand{\thelemma}{\Alph{section}\arabic{lemma}}

\section{Jordan Models} 

Let $\J$ be a euclidean Jordan algebra. As discussed earlier, 
this is associated with a probabilistic model $A(\J) = (X(\J),\M(\J),\Omega(\J))$, where $X(\J)$ is the set of 
primitive idempotents {\blue (that is, minimal projections)} in $\J$, $\M(\J)$ is the set of Jordan frames (maximal pairwise orthogonal sets of minimal projections), and 
$\Omega(\J)$ is the set of states on $(X(\J),\M(\J))$ arising from states on $\J$, that is, restrictions 
to $X(\J)$ of positive, normalized linear functionals on $\J$. 
%This is what is above called a {\em full} Jordan model.  
Using the self-duality of $\J$, it's easy to show that $\E(A) \simeq \J \simeq \V(A)$.  
%(More generally, a Jordan model is one in which 
%$\E(A)$ is a euclidean Jordan algebra and $\M(A)$ is a collection of Jordan frames. [AWK])

In this appendix, it is shown that any probabilistic model satisfying the conditions of Corollary 1 is 
actually a Jordan model of this form, and, conversely, that any such Jordan model satisfies the hypotheses of Corollary 2 (which imply those of Corollary 1).

\subsection{Direct Sums and central projections} 

At several points, we will need to use some basic facts about direct sum decompositions of Jordan algebras. 
{\blue The {\em direct sum} of Jordan algebras $\J_1,...,\J_n$ is the algebraic direct sum $\J = \J_1 \oplus \cdots \oplus \J_n$ of the vector spaces $\J_i$, consisting of $n$-tuples $(a_1,...,a_n)$ with $a_i \in \J_i$, with the Jordan product defined by 
$(a_i) \dot (b_i) = (a_i \dot_i b_i)$, 
where $\dot_i$ is the Jordan product on the $i$-th summand. Identifying $a \in \J_i$ with 
$(a_i)$ where $a_j = 0$ for $j \not = i$ and $a_j = a$ for $j = i$, we can treat each $\J_i$ as a subalgebra 
of $\J$, and write $(a_i) \in \J$ as 
$\sum_{i=1}^{n} a_i$. With this understood, we have $a \dot b = 0$ if $a \in \E_i$ and $b \in \E_j$ with $i \not = j$. The unit is evidently $\1 = \sum_{i=1}^{n} \1_i$ where $\1_i$ is the unit in $\J_i$. Note that the 
canonical projection map $\pi_i : \J \rightarrow \J_i$ is a Jordan homomorphism. Thus, 
$e \in \E$ is an idempotent iff $e_i := \pi(e)$ is an idempotent in $\J_i$. As 
$e = \sum_i e_i$, it follows that $e$ is a {\em primitive} idempotent iff $e = e_i \in \E_i$ for some $i = 1,...,n$. 
In other words, every  primitive idempotent of $\J$ lives in one of the summands $\J_i$.  

If each $\J_i$ is a euclidean Jordan algebra with inner product $\langle \ , \ \rangle_i$, then we endow 
$\J = \bigoplus_i \J_i$ with the usual inner product, that is, $\langle a, b \rangle \ = \ \sum_i \langle a_i , b_i \rangle_i$. 
where $a_i = \pi_i (a) $ and $b_i = \pi_i (b)$.  Thus, the summands $\J_i$ are orthogonal to one another as subspaces of $\J$. A euclidean Jordan algebra is {\em simple} iff it is not 
isomorphic to a direct sum of non-trivial Jordan algebras. Every EJA is isomorphic to a direct sum of simple EJAs in an essentially unique way. It will be helpful briefly to 
review how this decomposition works. }
%Let $\J$ be an EJA. 
Elements $a$ and $b$ in an EJA $\J$ {\em operator commute} with one another iff $a \dot (b \dot x) = b \dot (a \dot x)$ for 
every $x \in \J$, i.e., iff the operators $L_{a}$ and $L_{b}$ of left Jordan-multiplication by $a$ and $b$ commute. 
An element of $\J$ is {\em central} iff it operator-commutes with all elements of $\J$.  {\blue If $\J = \bigoplus_{i} \J_i$, then each of the units $\1_i \in \J_i$ is central. Note that these elements are also idempotents, or projections. }
%The set of all central elements, the {\em center} of $\J$, is 
%an associative Jordan sub-algebra, containing the unit element of $\J$ (\cite{Alfsen-Shultz} 1.52). 
A {\em Jordan ideal} of an EJA $\J$ is a subspace of the form $c\J$ where $c$ is a central projection, which is then unique. In this case, 
%$\J = c\J \oplus c'\J$, where $c' = \1 - c$, and 
the mapping $x \mapsto c\dot x$ is a Jordan homomorphism 
from $\J$ onto $c\J$; indeed, $x \mapsto (c\dot x, c' \dot x)$ provides a canonical isomorphism $\J \simeq c\J \oplus c' \J$. More generally, 
%any maximal collection of pairwise Jordan-orthogonal central projections $c_i$ provides an isomorphism 
%$\J \simeq \bigoplus_{i} c_i \J_i$, given by $x \mapsto (c_i \dot x)_{i}$ {\blue (\cite{Alfsen-Shultz}, 2.44)}. Thus, 
%an EJA is simple (or irreducible) 
%iff its center is trivial, i.e, it has no central projections other than $0$ or $\1$.   \\
{\blue recall that idempotents $e, f$ in a Jordan algebra are {\em Jordan-orthogonal} iff $e \dot f = 0$. 
%\noindent{\bf A.1 Proposition:} {\em 
If $\{c_i\}$ is a maximal pairwise Jordan-orthogonal set of central 
idempotents, then it is straightforward to show that 
$\sum_{i} c_i = \1$ and, hence, that $\J \simeq \bigoplus c_i \J$ via the 
mapping $\phi : a \mapsto (c_i a)$.   Moreover, 
each of the summands $\J_i := c_i \J$ is simple. For if  
$c_i = p + q$ where $p$ and $q$ are central projections in $c_i \J$, then %{\bf \blue [Need a lemma here] ADDRES!} 
$p, q$ are central in $\J$,  % [SAY WHY]
and are Jordan-orthogonal to every $c_j$ with $j \not = i$, so $\{c_i\}_{i \not = j} \cup \{p,q\}$ is a 
larger pairwise Jordan-orthogonal set of central projections, {\blue a} contradiction.
}

%[NOTE: This is the start of the proof of direct sum decomp... Do we need/want this?]
 %So each $c_i \J$ is simple. 
%{\blue [Ref 2 doesn't understand what is being proved by contradiction. Hopefully cleared up 
%by adding statement that each $c_i \J$ is simple, above.]}

\subsection{The unit effect as the Jordan unit}

Suppose $A$ is a model satisfying the hypotheses of Corollary 1.  %or Corollary 2. 
In particular, then, 
$A$ is HSD, so $\E := \E(A)$ has a Jordan structure. 
%Let $A$ be an HSD model, so that $\E := \E(A)$ has a Jordan structure. Assume also that every $x \in X(A)$ 
%has the same norm in $\E(A)$, as will be the case if the inner product on $\E$ arises from a weak conjugate. 
We wish to show that the unit effect $u \in \E(A)$ is (or can be taken to be) the Jordan unit, and {\blue that} each outcome $x \in X(A)$ --- or, more exactly, the corresponding 
effect $\hat{x}$ --- is a primitive idempotent. \footnote{This can actually be established very 
directly by appealing to a slightly stronger version of the Koecher-Vinberg Theorem, as in \cite{LTHSD}. The proof given here is more self-contained and elementary.} In fact, we will ultimately establish more, namely, that 
$X(A)$ corresponds exactly to the set of primitive idempotents, and $\M(A)$, to the set of Jordan frames, of $\E(A)$. 

%The inner product on $\E$ arising from the given weak conjugate gives all outcomes the same norm, namely 
%$1/\sqrt{n}$, where $n$ is the rank of $A$. 
In what follows, normalize the inner product on $\E$ so that $\langle x, x \rangle = 1$ for every $x \in X(A)$. 
(This is possible, since the inner product arising from a correlator assigns every outcome the same norm.)
Note that we also have $\langle u, x \rangle = 1$, and $\langle u, u \rangle = n$, the rank of $A$.  Every 
state $\alpha \in \V(A)$ corresponds to a unique $a \in \E(A)_+$ with $\alpha(x) = \langle a,x \rangle$. 
In particular, $\langle a, u \rangle = 1$. Conversely, every $a \in \E(A)_+$ with $\langle a, u \rangle = 1$ 
corresponds to a state in this way, since $\E(A) \simeq \V(A)$ as ordered spaces.  

The KV theorem produces a Jordan structure on $\E(A)$ in which the Jordan unit, $\1$, {\blue is fixed 
by every order-automorphism that is also an orthogonal transformation relative to the inner product. 
That is, writing $\Aut(\E)$ for the group of order-automorphisms and $\Aut(\E)_1$ for the stabilizer of 
$\1$ therein, $\Aut(\E) \cap O(\E) \subseteq \Aut(\E)_{\1}$. }
%{\blue [Ref 2: clarify 
%what is $u$ and what is $\1$]} is an element $\1 \in \E(A)$ such that
%\[\Aut(\E) \cap SO(\E) \subseteq \Aut(\E)_{\1},\]
%where $\Aut(\E)$ denotets the group of order-automorphisms of $\E$, {\blue and 
%$\Aut(\E)_{\1}$ is the stabilizer of $\1$ in $\Aut(\E)$.}
%That is, every orthogonal order-automorphism fixes $\1$ {\blue [as opposed to $u$...]}. 
Moreover, the stabilizer of $\1$ in the connected component $G$ of $\Aut(\E)$ is then exactly $K := G \cap SO(\E)$ \cite{FK}. 
 In {\blue particular,} 
every order-automorphism {\blue in the connected component of the identity} fixing $\1$ is an orthogonal transformation with respect to $\langle \ , \ \rangle$.\\

In the interest of notational simplicity, from this point 
on I will identify outcomes $x \in X(A)$ with the corresponding evaluation functionals $\hat{x} \in \E(A)$, treating  
each test $E \in \M(A)$ as a set of effects. \\

\noindent{\bf A.1 Lemma:} {\em For each $x \in X(A)$, there exists a primitive idempotent $e_x$ and a scalar $t_{x} > 0$ such that 
$x = t_{x} e_{x}$, and every primitive idempotent corresponds to an outcome in this way. }\\

\noindent{\em Proof:} Since $A$ is sharp and $\langle x, x \rangle = \langle x, u \rangle = 1$, we have 
$\delta_{x} = \langle x |$ for every $x \in X(A)$; that is, $\delta_{x}(a) = \langle x, a \rangle$ for 
all $a \in \E(A)$. In particular, as $\delta_x$ is a pure state, 
$x$ is ray-extremal for every $x \in X(A)$. Hence, $x = t_{x} e_{x}$ for some primitive idempotent $e_x$. Conversely, since $X(A)$ generates $\E(A)$, every ray-extremal element of $\E(A)_+$ must be a multiple of some $x \in X(A)$. In particular, then, each primitive idempotent is a  multiple of an outcome, and vice versa. $\Box$ \\

Notice that if $x$ and $y$ are distinct outcomes belonging to a common test, we have $\langle e_x, e_y \rangle = \frac{1}{t_x t_y} \langle x, y \rangle = 0$. 
Hence, $1  =  \langle x, x \rangle = t_{x}^{2} \|e_{x}\|^2$, 
so $t_{x} = 1/\|e_{x}\|$ for every $x$. \\

\noindent{\bf A.2 Theorem:} {\em The Jordan product on $\E(A)$ can be so chosen that $u_{\blue A}$ is the Jordan unit, {\blue $\1$} and every 
$x \in X(A)$ is a primitive idempotent.}\\

\noindent{\em Proof:} We  consider in turn the case in which $\E(A)$ is simple and the general case in which $\E(A)$ is a direct sum of simple ideals. Throughout, we write $u$ for $u_A$.

%\noindent 
{\em Case 1: $\E(A)$ is simple.}  By \cite{FK} Corollary IV.2.7, the group $K := G \cap SO(\E)$, where $G$ is the connected component of the identity in $\Aut(\E)$,  
acts transitively on the set of primitive idempontents. Since $K$ consists 
of orthogonal transformations, all primitive idempotents have the same norm. 
It follows that, for $\E$ simple,  $\|e\| \equiv c > 0$ for all primitive idempotents, whence, that 
we have $t_x \equiv t = 1/c$ for all $x$. That is, if $\E(A)$ is simple, $x = te_{x}$ for every $x \in X(A)$. 
%We now claim that then $u = t\1$.  By the spectral decomposition for states (equation (4)), $\1 = \sum_{x \in E} %s_{x} x = \sum_{x \in E} s_{x} t e_{x}$ for some $E \in \M(A)$. 
{\blue Now redefine the Jordan product on $\E$ by setting 
$a \circ b := t^{-1} a \dot b$. It is easy to check that this gives a Jordan product defining the same 
positive cone, with unit $\1' := t\1$. Also note that the new Jordan product continues 
to ineract with the given innner product in the desired way, i.e., $\langle a \circ b, c \rangle = \langle b, a \circ c \rangle$ for all $a,b,c \in \E$. We also have, for each outcome $x \in X(A)$, 
\[x \circ x = t^{-1}(t e_x \dot t e_x) = t(e_x \dot e_x) = 
t e_x = x,\]
so $x$ is an idempotent with respect to this new Jordan product; moreover, since the positive cone is unchanged, this is still ray-extremal, hence, a {\em primitive idempotent.} We now have 
\[\langle x, \1' \rangle = \langle x^2, \1' \rangle = \langle x, x \rangle = 1.\]
 By the spectral decomposition for states (equation (4)), we have 
$\1' =  \sum_{x \in E} s_{x} x$ for some test $E \in \M(A)$ and coefficients $s_x$. But for every $x \in E$, 
\[s_{x} = \sum_{y \in F} s_y \langle x, y \rangle = \langle x, \1' \rangle = 1, \]
so  $\1' = \sum_{x \in E} x = u_A$. } \\
 %Jordan-multiplying by any particular $x \in E$, we have [NEED $x \perp y$ implies $x \circ y = 0$! So better: 
%use $x \perp y \Rightarrow \langle x, y \rangle = 0$.] 
%\[x = \sum_{y \in E} t s_{y} x \circ y = t s_x x\]%
%whence, $t s_x = 1$ and $\1' = \sum_{x \in E} x = u$. }
\tempout{ Now for 
any $x\in E$, 
\begin{eqnarray*}
{\blue \1}  =  \langle x, x \rangle & = & \langle \1, x^2 \rangle \\
& = & \langle \sum_{y \in E} s_{y} t e_y, t^{2} e_{x} \rangle \\
& = & \sum_{y \in E} s_{y} t^{3} \langle e_y, e_x \rangle \\
& = & s_{x} t^{3} \|e_{x}\|^2 = s_{x} t.
\end{eqnarray*}
Hence, $\1 = \sum_{x \in E} e_{x}$.  
We now have $u = \sum_{x \in E} x = \sum_{x \in E} t e_{x} = t \1$.

We can now redefine the Jordan product by setting 
$a \dot b = \frac{1}{t} ab$. Then $u \dot a = \frac{1}{t} (t\1 a) = a$, so $u$ is the unit, and and $x \dot x = \frac{1}{t} (t_{x}^2 e_{x}) 
= t_{x} e_{x} = x$, so each $x \in X$ {\em is} a primitive idempotent. \\
%(NOTE that re-scaling preserves Jordan identity:
%\[ a \dot (a^2 \dot b) = \frac{1}{t^2} (a(a^2 b)) = \frac{1}{t^2} (a^2 (ab)) = \frac{1}{t}(a^2(\frac{1}{t} ab)) 
%= a^2 \dot (a \dot b).)\]
}
%\noindent 

{\em Case 2: $\E(A)$ a direct sum of simple ideals.} Let $\E = \bigoplus_{i=1}^{k} \E_i$ where {\blue each $\E_i$ is a simple Jordan algebra. By Lemma A.1, we still have a correspondence $x \mapsto e_x$ between 
outcomes $x \in X(A)$ and primitive idempotents $e_x \in \E$, with $x = t_{x} e_x$ for some $t_x > 0$. As remarked earlier, each primitive idempotent lies in a unique summand $\E_i$ whence, the same is true for outcomes. 
The argument given above shows that, for all $x \in \E_i$, we have $t_{x} = t_{i}$ for a constant $t_i > 0$ 
depending only on the summand. Adjusting the Jordan product on each summand as in Case 1, we obtain 
a new Jordan product on $\E$ given by 
\[(a_i) \circ (b_i) \ = \ \sum_{i} t_{i}^{-1} a_i \dot b_i\]
rendering each $x \in X(A) \cap \E_i$ --- and hence, each outcome $x \in X(A)$ --- a primitive idempotent.  By the 
same argument as in Case 1, we have $\langle x, \1 \rangle = 1$ for all $x \in X$. Expanding 
$\1 = \sum_{x \in E} s_x x$ for some test $E \in \M(A)$, we then have (again, just as in the irreducible case) 
that $1 = \langle x, \1 \rangle = \sum_{y \in E} s_y \rangle x, y \rangle = s_x$, 
whence, $s_x = 1$ for all $x \in E$, and hence, $\1 = u_A$. $\Box$ }\\

%[SHOW $x \perp y$ in $E$ means $x \circ y = 0$ ... ] 
%Moreover, if $\1' = \sum_{x \in E} s_x x$, we now have, by the same 
%argument as in Case 1, that $s_{x} = 1$ for every $x$, whence, $\1' = u$. }
%have $s_{x} x = x$, whence, .... }

\tempout{

 ideal with central projection $c_i$. 
Let $X_i$ and $P_i$ denote, respectively, the outcomes and primitive idempotents in $\E_i$, and note 
that $X(A) = \bigcup_{i} X_i$ and $P(\E) = \bigcup_i P_i$. The correspondence $x = t_{x} e_{x}$ 
sets up a bijection between $X_i$ and $P_i$, and, as above, $t_{x} = 1/\|e_{x}\|$. Since the 
goup $K_i = G_i \cap SO(\E_i)$ acts transitively on $P_i$, we have $t_{x} \equiv t_i$ for all $x \in X_i$.

Let $u_i = c_i u$ and $\1_i = c_i \1$ (the latter 
being the Jordan unit in $\E_i$). For every $x \in X_i$ we have 
\[\langle u_i, x \rangle = \langle c_i u , c_i x \rangle = \langle u, c_{i}^{2} x_i \rangle  = \langle u, x \rangle = 1.\]
%As we also have $\langle x, x \rangle = 1$ for all $x \in X_i$, 
Now let $\1 = \sum_{x \in E} s_x x$ for some $E \in \M(A)$, as above. Letting $E_i = E \cap X_i$, we have 
\[\1_i = c_i \1 = \sum_{x \in E} s_{x} c_{i}x = \sum_{x \in E_i} s_{x} x.\]
Repeating the argument above for each simple summand $\E_i$, we find that $\1_i = \sum_{x \in E_i} e_{x}$. 
It follows that $\1 = \sum_{x \in E} e_{x}$. We also have 
\[u_i = c_iu = \sum_{x \in E} c_i x = \sum_{x \in E_i} x = t_i \left (\sum_{x \in E_i} e_{x} \right ) = t_i \1_i.\]
Thus, 
\[\sum_{i} t_{i} \1_i = \sum_i u_i = u = t\1 = \sum_i t\1_i.\]
Taking the Jordan product on both sides with $c_i$, we see that $t_i \1_i = t\1_i$, i.e., $t_i \equiv t$. Redefining the Jordan product on each simple summand $\E_i$, as above, we can take  
$t_i = 1$ for each $i$, and the proof is complete. $\Box$ \\
}

From now on, we treat $u_A$ as the Jordan unit of $\E(A)$ without further comment, and revert to writing 
$u$, rather than $\1$, for the unit in an abstract Jordan algebra.  It will be convenient at this point to 
adopt the notation $x \perp y$ to indicate that two outcomes $x, y \in X(A)$ are distinct and belong to a common 
test.\\
% Note that this does not (yet) imply that $x$ and $y$ are orthogonal with respect to the inner product 
%on $\E(A)$.\\

{\blue \noindent{\bf A.3 Corollary:}  {\em Let $x, y \in X(A)$. Then $x \perp y \Rightarrow x \dot y = 0$.} \\

\noindent{\em Proof:} By \cite{Alfsen-Shultz}, Prop. 2.18, if $p, q$ are projections in an EJA, then 
$p \leq u - q$ implies $p \dot q = 0$. By Proposition A.2, $x$ and $y$ are projections. 
If $x \perp y$, then $x + y \leq u$, so $x \leq u - y$. $\Box$. \\

Idempotents $p, q$ in a Jordan algebra are {\em Jordan-orthogonal} iff $p \dot q = 0$. Recall that a {\em Jordan frame} in an EJA $\E$ is a set $E$ of non-zero Jordan-orthogonal primitive idempotents summing to the unit. 
It follows from Proposition A.2 and Corollary A.3 that every test in $\M(A)$ defines a Jordan frame in $\E(A)$. We now show that, conversely, 
every Jordan frame in $\E(A)$ belongs to $\M(A)$.  \\}
%the models arising from our assumptions contain {\em all} Jordan frames.\\

\noindent{\bf A.4 Theorem:} {\em Let $A$ be any probabilistic model satifsying the hypotheses of Corollary 1 or of Corollary 2, and let $\E(A)$ have the Jordan structure in which $u_A$ is the Jordan unit, as per Lemma A.2. Then every Jordan frame of $\E(A)$ corresponds to a test in $\M(A)$. Hence, $A$ is a Jordan model. }\\ 

\noindent{\em Proof:} Theorem  A.2 and Corollary A.3 tell us that the set $X(A)$ of outcomes of $A$ is exactly the set of primitive idempotents (still continuing to identify outcomes with the corresponding effects), and that every test $E \in \M(A)$ is a Jordan frame of $\E(A)$. The Spectral Theorem for euclidean Jordan algebras (\cite{FK} Theorem III.1.1) tells us that if $\J$ is any EJA with 
unit $u$ and $a \in \J$, then there exists a {\em unique} 
family of mutually {\blue Jordan-orthogonal} non-zero idempotents $e_i$ %with $\sum_i e_i = u$ 
and {\em distinct} coefficients $s_i$, such that $a = \sum_{i} s_i e_i$. 
{\blue Suppose the idempotents $e_i$ are primitive, and that $a$ can also be expanded as $\sum_{j} t_j f_j$ for 
some pairwise Jordan-orthogonal non-zero idempotents $f_j$ and coefficients $t_j$ (not a priori distinct). Then by collecting those terms with common coefficients, we can also write 
$a = \sum_{k} k p_{k}$ where $p_k = \sum \{ f_{j} | t_j = k \}$. Since the $f_j$ are pairwise Jordan-orthogonal idempotents,
it is easy to check that $p_k$ is also an idempotent, with $p_k \dot p_{k'} = 0$ for $k \not = k'$. It 
follows from the uniqueness of the spectral expansion that, for each $k$, there is some $i$ with $k = s_i$ 
and $p_k = e_i$. Since $e_i$ is primitive, this means that the sum $p_k = \sum\{f_j : t_j = k\}$ has 
a unique term, %has a unique term, 
call it $f_{j_i}$, with $f_{j_i} = e_i$. It follows that the coefficients $t_j$ were after all 
distinct, and the projections $f_j$ coincide with the projections $e_i$. In other words, if $a$ can be expanded with {\em dinstinct} coefficients relative to {\em some} family of mutually orthogonal primitive idempotents, it can have no other 
such expansion in terms of mutually orthogonal idempotents, with {\em any} coefficients.  Now let $F = \{e_1,...,e_n\}$ be a Jordan frame of $\E(A)$. Choosing distinct coefficients $s_i$, $i = 1,...,n$, let  $a = \sum_{i} t_i e_i$.  By our spectral decomposition for HSD models (equation (4)), there exists a test $E \in \M(A)$ and coefficients $t_x, x \in E$ such that $a = \sum_{x \in E} t_{x} x$.  Since $E$ is a Jordan frame,  
the uniqueness result above tells us that $E = F$. Thus, every Jordan frame of $\E(A)$ is a test in $\M(A)$, as 
advertised. $\Box$ }

\subsection{Conjugates and Filters for Jordan Models}

As discussed in Section 5, the question of what it means, physically, to say that a given system 
has a conjugate really depends on the probabilistic theory --- and the notion of composite system --- 
with which one is concerned. But if we are content to interpret this idea very broadly, we can understand the composite $A \bar{A}$ as the 
``maximal" tensor product $A \otimes_{\mbox{max}} \bar{A}$ \cite{BW12}, the states of which are simply the positive, normalized bilinear forms on $\E(A) \times \E(\bar{A})$. 
In particular, if $A = A(\J)$ is the model associated with a euclidean Jordan algebra $\J$, the canonical inner product on $\J$, normalized so that $\langle u, u \rangle = 1$,  
supplies exactly the needed bilinear form.  
Thus, every Jordan model can be regarded as its own weak conjugate. (At least, this is so if we construe ``$A$ has a weak conjugate" to mean only that 
there exists a perfectly correlating positive, bilinear form on $\E(A)$ --- mathematically, a {\em weaker} condition than that $A$ have a conjugate in any particular probabilistic theory.)

We now show that Jordan models support arbitrary filters, and that every state of such a model can be prepared 
by a symmetric filter.
For any element $a$ of a euclidean Jordan algebra $\J$,  the  operator $U_a : \J \rightarrow \J$  defined by 
\[U_a(x) := 2a\cdot(a \cdot x) - a^2 \cdot x\]
 is positive (\cite{Alfsen-Shultz} Theorem 1.25). If $\J$ is special, i.e., if $\J$ consists of self-adjoint 
 operators on a real, complex or quaternionic Hilbert space $\H$, with $a \cdot b = \tfrac{1}{2}(ab + ba)$, 
 one can check that $U_{a}(x) = axa$.  %Note that $U_{\sqrt{a}} u = a$. 
 Below, we shall see that if $a = \sum_{x \in E} t_{x} x$, where $E$ is a 
 Jordan frame, then $U_{a}$ is a filter with coefficients $t_x$.
 
Let $\J$ be an EJA. An element $a \in \J$ is {\em invertible} iff there exists an element $b$ of the associative 
sub-algebra  generated by $a$ and $u$, such that $a \dot b = u$. This element, which is clearly unique, is 
then the {\em inverse} of $a$, denoted by $a^{-1}$. The following collects some facts about invertibility that 
will be needed in a number of places below.\\

\noindent{\bf A.5 Proposition:} {\em The following are equivalent:
\begin{mlist} 
\item[(a)] $a$ is invertible;
\item[(b)] There exists some $b \in \J$ with $a \dot b = u$ and $a^2 \dot b = a$;
\item[(c)] $U_{a}$ is invertible. In this case, $U_{a}^{-1} = U_{a^{-1}}$.\\
\end{mlist}}

\noindent{\em Proof:} That (a) implies (b) is clear. That (b) implies (a) is a consequence of the fact (the Shirsov-Cohn Theorem; 
see \cite{Alfsen-Shultz}, Proposition 1.14) that the Jordan algebra generated by two elements and $u$ is special, plus the fact 
that (a) and (b) are equivalent in special Jordan algebras. See \cite{Alfsen-Shultz} Lemma 1.16 and Proposition 1.17 for details.
The equivalence of (a) and (c) is Lemma 1.23 in \cite{Alfsen-Shultz}. $\Box$ \\

We shall also need the following elementary observation:\\

\noindent{\bf A.6 Lemma:} {\em If $a$ has a spectral decomposition $a = \sum_{x \in E} t_x x$ where $E$ is a 
Jordan frame and $t_x \not = 0$ for all $x \in E$, then $a$ is invertible. In particular, $a$ is invertible if lies in the interior of $\J_{+}$. } \\

\noindent{\em Proof:} By spectral theory, $a = \sum_{x \in E} t_x x$ where $E$ is a Jordan frame. If $t_x \not = 0$ for all $x$, let $b = \sum_{x \in E} t_{x}^{-1} x = f(a)$ where $f(x) = x^{-1}$, so that 
$b \in C(a,u)$, the Jordan subalgebra generated by $a$ and $u$. Observe that, by the Jordan-orthogonality of elements of a Jordan frame, %[CHECK -- note part of def. of Jordan frame in FK] we have 
\[a \dot b = \sum_{x,y \in E} t_{x}t_{y}^{-1} x \dot y = \sum_{x} t_{x} t_{x}^{-1} x = \sum_{x \in E} x = u.\]
Thus $a$ is invertible with inverse $b$.
 $\Box$ \\ %[AWK]\\

\noindent{\bf A.7 Lemma:} {\em Let $\J$ be any EJA. Every state of the model $A(\J)$ is preparable by symmetric filter, and every non-singular state, by a reversible symmetric filter.}\\

\noindent{\em Proof:}  Let $\alpha$ be any state on $\J$. By self-duality, there exists a unique $w \in \J_+$ with 
$\alpha(b) = \langle w, b \rangle$ for all $b \in \J$. The spectral theorem gives us a Jordan frame $E$ and 
a decomposition $w = \sum_{x \in E} t_x x$. Let $a = \sum_{x \in E} t^{1/2}_{x} x \in \J_+$: then we have 
$U_a (u) = w$ and $U_a(x) = t_x x$ for every $x \in E$, so $U_a$ is a filter. Since left multiplication by $a$ is self-adjoint with respect to the 
inner product on $\J$, so is $U_a$, whence, $U_a$ is a symmetric filter. Finally, if $w$ lies in the interior of $\J_+$,  then the coefficients $t_x$, and hence also $t_{x}^{1/2}$, are all strictly positive. Thus, $a$ is invertible by Lemma A.5, and thus $U_{a}$ is invertible with inverse 
$U_{a^{-1}}$, again a positive operator, by Proposition A.5.  $\Box$ \\

As noted in the discussion preceding Lemma 5, every state preparable by a symmetric reversible filter is the marginal of a correlating state. Hence, every Jordan model is its own 
(strong) conjugate. Since such a model is evidently sharp, the conditions of Corollary 1 are satisfied. This proves Theorem 2: every probabilistic model satisfying the hypotheses of 
Corollary 1 or Corollary 2 is a Jordan model, and every Jordan model satisfies the hypotheses of both Corollaries.

\section{Symmetry and Subspace Axioms} 

\setcounter{lemma}{0}
  \renewcommand{\thetheorem}{\Alph{section}\arabic{theorem}}

In most other recent reconstructions of QM \cite{Hardy, Dakic-Brukner, Masanes-Mueller, CDP}, one encounters some version of a {\em subspace axiom}. Informally, the idea is that if we {\em constrain} the states of a given system so as to render 
a certain measurement result impossible, we obtain what amounts the state space of a 
new system in its own right, to which any other axioms must then apply. 
In practice, the subsets of the state space arising in this way are (some of the) {\em faces} of the 
larger state space\footnote{Recall that a face of a convex set $K$ is a convex set $F \subseteq K$ such 
that for all $\alpha, \beta \in K$ and all $t \in (0,1)$, $t\alpha + (1 - t)\beta \in F$ implies $\alpha, \beta \in F$.}.  
An assumption of this sort was made by Hardy \cite{Hardy}, and, 
following his lead, 
is also central 
in reconstructions by Dakic and Brukner \cite{Dakic-Brukner}, Masanes and Mueller \cite{Masanes-Mueller} and Chiribella, D'Ariano and Perinotti \cite{CDP},  
although the precise formulation varies somewhat from one set of authors to another.

In this Appendix it is established that, for uniform probabilistic models, a certain version of this subspace axiom, 
plus a very reasonable compactness assumption, enforce a spectral decomposition of states (equation {\blue (4)} in Section 3). Thus, these assumptions can replace condition (b) in the definition of a conjugate, {\blue at least 
as far as Theorem 1 is concerned.}  %{\blue [Ref 2: ``only as far as Theorem 1 is concerned". CHECK!!]}
% and thus, provide an alternative route to self-duality.  
Moreover, a stronger version of the 
subspace axiom, which seems about equally well motivated,  implies 
the existence of arbitrary reversible filters. Combined with a rather weak and operationally natural symmetry assumption, 
this in turn yields the homogeneity of the space $\V(A)$. 
In the presence of weak conjugates, it follows that $\V(A)$ is both homogeneous and self-dual.\\

\noindent{\bf Probabilistic Theories} Before proceeding, it will be important to clarify the term ``probabilistic theory", up to this point used rather freely and informally.  In the very broadest sense, a probabilistic theory is simply a class of probabilistic models, together with 
some designated processes that are singled out for study. But allowing for the composition of processes, and assuming that identity operators count as processes, it is very natural to assume that a probabilistic theory 
is a {\em category} of probabilistic models and processes.  This is the point of view taken 
in the ``categorical quantum mechanics" programme of Abramsky, Coecke and others \cite{Abramsky-Coecke}. 
It is also usual in this context also to assume that this category has a (symmetric) monoidal structure, 
allowing for the formation of composite systems. However, for my purposes, it will 
be enough to require only that we are given a class $\Cat$ of probabilistic models and, for each 
model $A \in \Cat$, a preferred monoid $\Proc(A)$ of allowed processes, i.e., positive linear mappings 
$T : \V(A) \rightarrow \V(A)$ with $u_{A}(T(\alpha)) \leq 1$ for all $\alpha \in \Omega(A)$.\footnote{We can, 
of course, regard this as a degenerate category in which there are no morphisms between different 
objects. My intention, however, is simply to leave it open which mappings between different 
systems count as processes, as nothing will depend on this. This is another respect in which the 
approach taken here is ``single-system".}
%A process $T \in \Proc(a)$ is {\em p-reversible} iff $T$ is invertible, $T^{-1}$ is positive, and, for 
%some $p \in [0,1]$, $pT^{-1} \in \Proc(A)$. In this case, we regard $p$ as the probability with which $T$ can 
%be reversed. If $T$ is p-reversible with $p=1$, I shall simply say that $T$ is {\em reversible}. This 
%is equivalent to the condition that $T^{\ast} u_{A} = u_{A}$. 

We continue to identify outcomes with the corresponding effects, regarding $X(A)$ as a subset of $\E(A)$ and 
$\M(A)$, as a collection of subsets of $\E(A)$.  
We can then define a {\em symmetry} of $A$ to be 
a dual process $g = \phi^{\ast}$ where $\phi \in \Proc(A)$ is invertible, such that $g \M(A) = \M(A)$. 
I will write $G(A)$ for the set of all symmetries of $A$, noting that this is a group. As above, 
we use the notation $x \perp y$ to indicate that $x$ and $y$ are distinct outcomes 
belonging to a common test. Notice that this does not (yet) imply that $x$ and $y$ are orthogonal with respect 
to any inner product on $\E(A)$.

\subsection{The Subspace Axiom and Spectrality}

{\bf Masanes and Mueller's Framework} In the reconstruction due Masanes and Mueller \cite{Masanes-Mueller}, one begins 
by specifying the state-space of a physical system, taken to be a finite-dimensional compact convex set $\Omega$. 
Measurement results are associated with {\em effects}, i.e., affine functionals $a : \Omega \rightarrow \R$ 
with $0 \leq a(\alpha) \leq 1$ for all $\alpha \in \Omega$. Of course, $a(\alpha)$ is understood as the probability 
of the result associated with $a$ being obtained, when the system's state is $\alpha$. Masanes and Mueller do {\em not} assume 
that all effects correspond to physically realizable measurement results, but {\em do} seem, tacitly, 
to define a {\em measurement} to be {\em any} list $a_1,...,a_n$ of allowed effects that sum to the 
unit effect $u$ (where $u(\alpha) = 1$ for all $\alpha \in \Omega(A)$). Further assumptions, 
explicit in \cite{Masanes-Mueller2}, are that the set of allowed effects is topologically closed, convex, and 
closed under $a \mapsto u - a$. 
Masanes and Mueller call states $\alpha_1,...,\alpha_n \in \Omega(A)$  {\em perfectly distinguishable} 
iff there exist affine functionals  $a_i : \Omega \rightarrow \R$ with $0 \leq a_i(\alpha) \leq 1$, representing 
measurement results, such that $a_1 + \cdots + a_n = u$, where $u$ is the unit effect ($u(\alpha) = 1$ for all $\alpha \in \Omega$) and 
$a_i (\alpha_j) = \delta_{i,j}$ for all $i, j = 1,...,n$. The {\em information capacity} of the system is 
the maximum size of such a perfectly distinguishable set of states. A {\em complete measurement} is 
a measurement that perfectly distinguishes a maximum number of states. In the actual development of 
their results, they effectively take $\M(A)$ to be the set of complete measurements in this sense. \\

\noindent{\bf Masanes and Mueller's Subspace Axiom} Here is how Masanes and Mueller explain their version of the subspace axiom in \cite{Masanes-Mueller2}:
\begin{quote} 
{\bf Postulate 2 (Equivalence of subspaces).} {\em Let $S_N$ and
$S_{N-1}$ be systems with capacities $N$ and $N-1$, respectively.
If $E_1, . . . , E_N$ is a complete measurement on $S_N$ ,
then the set of states $\omega  \in S_N$ with $E_N(\omega) = 0$ is equivalent
to $S_{N-1}$.}

The notion of equivalence needs some discussion. Postulate
2 states the equivalence of $S_{N-1}$ and
\[S_{N-1}'
:= \{\omega \in S_N | E_N (\omega) = 0\}.\]
Denote the real linear space which contains $S_N$ by $V_N$;
define $V_{N-1}$ analogously, and set 
$V_{N-1}' 
:=\spn(S_{N-1}')$.
Equivalence means first of all that there is an invertible
linear map $L : V_{N-1} \rightarrow V_{N-1}'$
such that $L(S_{N-1}) = S_{N-1}'$. But it also means that transformations 
... 
% and measurements
on one of them can be implemented on the
other... %The transformations on $S_{N−1}'$ are defined analogously.
To be more specific, define $\bar{G}_{N-1}'$
as the set of transformations
in $S_N$ that preserve $S_{N-1}'$ ...
\[\bar{G}_{N-1}' \ := \ \{T \in G_{N} \ |\  T S_{N-1}' = S_{N-1}'\}.\]
The set of reversible transformations 
$G_{N-1}'$ is defined as the restriction of all these transformations to $S_{N-1}'$...
\end{quote}
%{\bf [Extension property explained further on. But does it imply $gx = x$??]}

\noindent{\bf A Reformulation} Returning to our own formalism, 
suppose that $x \in X(A)$, and let 
\[F_{x} := \{~ \alpha \in \Omega(A) \ |\  \alpha(x) = 0 ~\}.\] 
Note that this is a face 
of $\Omega(A)$, corresponding exactly to the restricted state space contemplated in Masanes and Mueller's subspace axiom. If we wish to treat this as the state space of a model in our sense, we must associate a test space 
with it. The simplest option is to define, {\blue for $x \in X$,}
\[\M(A)_{x} = \{ \ E \setminus \{x\}  ~|~  E \in \M(A) \ \mbox{with} \ x \in E \}.\] 
Notice that {\blue the outcome-space of $\M(A)_{x}$ is the set of outcomes in $X(A)$ that are distinguishable 
from $x$; that is, the union of the tests in $\M(A)_{x}$ is the  set $\{\ y \in X(A) \ |\ y \perp x\ \}$. }
In the special case of a quantum model, say $A = A(\H)$, this is the {\em right} definition, since if $\K = x^{\perp}$ is the subspace of $\H$ orthogonal to a unit vector $x$, then any frame $F \in \M(\K)$ extends to a 
frame $E = F \cup \{x\}$ of $\H$, so that $\M(\K) = \M(\H)_{x}$.  

Any state $\alpha \in F_{x}$ defines a probability weight $\alpha_{x}$ on $\M(A)_{x}$, simply by restricting 
$\alpha$ to outcomes in $\bigcup \M(A)_x$, i.e., to outcomes $y \in X(A)$ with $y \perp x$.  The mapping 
$\alpha \mapsto \alpha_x$ is obviously affine, but in general is not injective. For a simple example, consider 
the ``square bit": 
$\M(A) = \{\{x, x'\}, \{y, y'\}\}$, with $\Omega(A)$ the set of all probability weights thereon. Then 
$\Omega(A)$ is isomorphic to the unit square in $\R^2$ under the mapping $\alpha \mapsto (\alpha(x), \alpha(y))$. 
The face $F_{x}$ can be identified with the right-hand face of the square, i.e., $F_{x} = \{ (0,t) | 0 \leq t \leq 1\}$.
On the other hand, $\M(A)_{x}$ is the trivial test space $\{ \{x'\} \}$, which has only a single probabilty weight. 
We will therefore need to assume that $\M(A)_{x}$ is large enough to separate points of $F_{x}$, i.e., 
that if $\alpha, \beta \in F_{x}$ are distinct, then there exists some $y \perp x$ with 
$\alpha(y) \not = \beta(y)$. This 
makes $\alpha \mapsto \alpha_x$ an injection, so that we can identify $F_{x}$ with a set of probability weights 
on $\M(A)_{x}$. \\

\tempout{
\noindent{\bf Definition B.1} A probabilistic model $A$ {\em has subspaces} iff, for every outcome 
$x \in X(A)$, the set of outcomes $\{ y \in X(A) | y \perp x\}$ separates points of $F_{x}$. Equivalently, 
the mapping $\alpha \mapsto \alpha_{x}$ is injective. When this is the case, we identify 
$F_{x}$ with its image under this mapping, and define the {\em submodel} corresponding to $x$ to be 
\[A_{x} \ := \ (\M(A)_{x}, F_{x}).\]
}

%\noindent{\bf Lemma B.2} {\em A probabilistic model that has subspaces, is sharp.}

%{\em Proof:} By induction on rank. Clearly, all models of rank $1$ are sharp. Assume the result holds 
%for all models of rank $n$ or lower. If $A$ has rank $n+1$ (with $n > 0$), and $y \in X(A)$, then since 
%$n + 1 > 1$ we can find some $x \perp y$. Since $
 
With this in mind, the following now seems to capture the spirit of Masanes and Mueller's axiom:\\

\noindent{\bf Definition B.1}  Say that a probabilistic theory $\Cat$ has the {\em subspace property} iff, for every 
$A \in \Cat$ and every $x \in X(A)$, 
\begin{mlist} 
\item[(i)]  $\M(A)_x$ separates points of $F_x$,  
\item[(ii)] The model $A_{x} := (\M(A)_{x}, F_{x})$ belongs to $\Cat$, and 
\item[(iii)] every symmetry in $G(A_x)$ extends to a symmetry $g \in G(A)$ with $gx = x$.\\
\end{mlist} 

As an example, let $A = A(\H)$ be the quantum model associated with a Hilbert space $\H$. If 
$x \in X(\H)$ is a unit vector, then $F_{x}$ is the set of density operators $W$ such that 
$\Tr(Wx) = 0$, or, equivalently, such that $P W P = W$ where $P$ is the orthogonal projection 
onto the subspace $\K_{x} = x^{\perp}$ orthogonal to $x$. In this case $\M(A)_{x}$ is the set of orthonormal bases 
of $\K$, so $\M(A)_{x} = \M(\K)$, and $A(\H)_{x} = A(\K_{x})$. \\

%{\em For the balance of this section, $\Cat$ will denote a probabilistic theory with the subspace property, 
%in which all models are finite-dimensional and uniform.}

\noindent{\bf Lemma B.2} {\em Let $\Cat$ be a probabilistic theory with the subspace property, in which 
every model is uniform. Then every model in $\Cat$ is sharp.} \\

\noindent{\em Proof:} By induction on rank. Clearly, all models of rank $1$ are sharp. Assume the result holds 
for all models of rank $n$ or lower. Suppose $A$ has rank $n+1$ (with $n > 0$), and let $x \in X(A)$. 
Since $n+1 > 1$, we can find some $y \perp x$. Since $A_{y} \in \Cat$ has rank $n$, $A_{y}$ is sharp. 
Therefore, there exists a unique $\delta_x \in F_{y}$ --- and hence, a unique $\delta_x \in \Omega(A)$ --- 
with $\delta_x(x) = 1$. $\Box$ \\

\noindent{\em Remark:} This is the only use we make of Condition (i) in Definition B.1. If one is content 
to assume that every model in $\Cat$ is sharp, this condition can be dispensed with.\\

When dealing with finite-dimensional probabilistic models, one usually assumes that the 
state-space $\Omega(A)$ is compact. It is equally natural to suppose that $X(A)$ is compact 
in the topology inherited from $\E(A)$ --- equivalently, in, the coarsest topology making 
every state $\alpha : X(A) \rightarrow [0,1]$ continuous. This is 
certainly the case for quantum models, where $X(A)$ is the unit sphere in a 
finite-dimensional Hilbert space, and can be shown to hold for the test space of 
``complete measurements" considered (implicitly) in \cite{Masanes-Mueller}. %\\
%We will simply say that $\Cat$ has {\em compact outcome-spaces} 
%More broadly if $A$ has 
%a compact group of symmetries acting transitively on $X(A)$, the latter will carry a natural 
%compact topology.\\

Let us say that a probabilistic model $A$ is {\em spectral}  if it is sharp, and every states $\alpha \in \Omega(A)$ has a spectral decomposition $\alpha = \sum_{x \in E} \alpha(x) \delta_{x}$ for some $E \in \M(A)$. \\

\noindent{\bf Proposition B.3} {\em Let $\Cat$ be a probabilistic theory with the subspace property, in which 
all models are uniform and have compact outcome-spaces. Then every model in $A$ is spectral.}\\ 
%%% <--- HAVE WE DEFINED "SPECTRAL"?? % then $A$ is spectral.} 

\noindent{\em Proof:} By induction on the rank of $A \in \Cat$. Spectrality is trivial for 
models of rank $1$ (which have only a single state). Assume the result holds for all $A \in \Cat$ having rank $< n = \mbox{rank}(A)$. Let $\alpha \in \Omega(A)$. Since $X(A)$ is compact and $\alpha$ is continuous on $X(A)$, $\alpha$ takes its minimum 
value, $m$, $0 \leq m < 1$, at some point $x_o \in X(A)$. Thus, $\beta := \alpha - m \delta_{x_o}$ is non-negative on $X(A)$, hence, belongs to $\V(A)_+$. Now, $u_A(\beta) = 1 - m$, so $\beta_{1} := (1 - m)^{-1} \beta \in \Omega(A)$, 
and $\alpha = (1 - m)\beta_1 + m \delta_{x_o}$.  
Since  
$\beta_{1}(x_o) = 0$,  $\beta_1$ belongs to the face $F_{x_o} := \{ \alpha \in \Omega(A) | \alpha(x_o) = 0\}$. 
Because $\C$ has the subspace property, $A_{x_o}$ belongs to $\Cat$. Since $A_{x_o}$ has rank $n -1$, our 
inductive hypothesis implies that $\beta_1 = \sum_{y \in F} \beta_1(y) \delta_y$ for some $F \in \M(A_{x}) 
= \M(A)_{x}$, i.e., for some $F = E \setminus \{x_o\}$, $x_o \in E \in \M(A)$. But now 
\[\alpha = (1 - m)\beta_1 + m \delta_{x_o} = \sum_{y \in F} (1 - m) \beta(y)\delta_{y} + m \delta_{x_o}\]
which gives a spectral decomposition for $\alpha$. $\Box$ \\

\noindent{\em Remarks:}
 %(1) 
Subspace axioms are usually paired with a conceptually distinct requirement that all systems of a given information capacity are isomorphic. This immediately rules out any {\blue non-classical} theory involving superselection rules, or in which there are more than one kind of ``bit".  As we do not impose such an isomorphism requirement, we avoid this restriction.

\subsection{Subspaces plus symmetry}  

%[THIS SECTION SEEMS TO DEPEND ON OLDER DEF OF $\M_x$! WITH NEWER, FULL SYMMETRY SHOULD BE EASY!]
In addition to one or another version of subspace postulate, \cite{Hardy, Dakic-Brukner, Masanes-Mueller, CDP} also 
 assume that every model $A$ to carries a preferred 
group $G(A)$ of symmetries, understood as acting on the state space, which is consistent with our choice of 
$G(A)$ in the previous section. 
The first three of the cited papers assume a {\em symmetry postulate} requiring that $G(A)$ act 
transitively on pure states, i.e., on the extreme points of $\Omega(A)$. This is also a consequence of the 
``purification postulate" used in \cite{CDP}.  \\

\noindent{\bf Definition B.4} A probabilistic model $A$ is {\em symmetric}
%\footnote{\bf NOTE: this differs from 
%previous usage, as I don't require trans. on $\M(A)$...}  
iff $G(A)$ acts transitively on $X(A)$. If $A$ is sharp 
and  all pure states are of the form $\{\delta_x | x \in X(A)\}$,  this implies $G(A)$ also acts transitively on pure states. \\

As an example, let $A = A(\H)$ be the quantum model associated with a Hilbert space $\H$. 
As discussed earlier, this means that $\M(A)$ is the set of orthonormal bases of $\H$, $X(A)$ is (therefore) 
the unit sphere of $\H$, and $\Omega(A)$ is 
the set of density operators on $\H$. If we take $G(A)$ to be the group of unitary operators on $\H$, then 
$G(A)$ certainly acts transitively on $X(A)$. 

Notice that, in this example, $G(A)$ acts transitively also 
on $\M(A)$. In \cite{Wilce09} and elsewhere, a uniform test space with a preferred group of symmetries 
$G(A)$ is said to be {\em fully symmetric} if, for every pair of tests $E, F \in \M(A)$ and every bijection 
$f : E \rightarrow F$, there exists an element $g \in G(A)$ such that $gx = f(x)$ for every $x \in E$. 
If $A$ has rank $n$, we can define an {\em ordered test} to be an $n$-tuple $(x_1,...,x_n) \in X^{n}$ 
where $\{x_1,...,x_n\} \in \M(A)$. Then full symmetry is the condition that $G(A)$ act transitively 
on ordered tests, where $g(x_1,...,x_n) = (g x_1,...,g x_n)$. \\

%\newpage
\noindent{\bf Lemma B.5} {\em Suppose that $\Cat$ has the subspace property. If every $A \in \Cat$ is uniform and symmetric, then every $A \in \Cat$ is fully symmetric.}\\
 
\noindent{\em Proof:} Every symmetric model of rank $1$ is trivially fully symmetric. Suppose every model 
in $\Cat$ having rank $\leq n$ is fully symmetric, and let $A \in \Cat$ be a model of rank $n + 1$. 
Let $(x_o,...,x_n)$ and $(y_o,...y_n)$ be any ordered tests of $A$. By symmetry, we can find 
some $g_o \in G(A)$ such that $g_o x_o = y_o$. Now $(y_1,...,y_n)$ and $(g_o x_1,...., g_o x_n)$ are both 
ordered tests of $A_{y_o}$. By our induction hypothesis, $A_{y_o}$ is fully symmetric, so there exists 
some $g_1 \in G(A_{y_o})$ with $y_i = g_{1} g_o x_i$ for every $i = 1,....,n$. By the subspace property, 
$g_1$ extends to a symmetry $g \in G(A)$ with $g y_o = y_o$ and $g z = g_{1} z$ for every $z \perp y_o$ --- 
in particular, $g g_o x_i = g_1 g_o x_i = y_i$ for $i = 1,....,n$. Hence, $g g_{o}$ takes $(x_o,....,x_n)$ to 
$(y_o,...,y_n)$. $\Box$ 

%It is important to note that symmetry restricts us to {\em indecomposable} models.  Given sets $X$ and $Y$, I will write $X \oplus Y$ for their coproduct. Given probabilistic models $A$ and $B$, define $A \oplus B$ to be the model with  test space %$\M(A \oplus B) = \{ E \oplus F | E \in \M(A), F \in \M(B)\}$, whence, outcome-set $X(A \oplus B) = X(A) \oplus X(B)$, and with state-space $\Omega(A) \oplus \Omega(B) := \{ \alpha : X(A) \oplus X(B) \rightarrow \R  | 
%\alpha|_{X} \in \Omega(A) \ \mbox{and} \ \alpha|_{Y} \in \Omega(B)\}$.   HM.  May need to emulate situation with $G(A \oplus B) = G(A) \times G(B)$?? But can permute factors if isomorphic....

\subsection{A Strengthened Subspace Postulate}

\tempout{Let $A$ be sharp. If $E \in \M(A)$, let $\Delta(E) \subseteq \Omega(A)$ be the closed convex hull of the pure states $\delta_x$, $x \in E$. Note that this is a simplex. Write $\V(E)$ for the span in $\R^{X(A)}$ of $\Delta(E)$, 
ordered by the cone having $\Delta(E)$ as a base, so that $\V(E) \leq \V(A)$ as a model. (Notice, however, 
that $\Delta(E)$ is not a face of $\Omega(A)$, in general, so, does not correspond to a subspace in 
the sense of Definition B.1 above.) 

The idea is that any process $\phi_{E} : \V(E) \rightarrow \V(F)$ that would be allowed by classical probability theory,  
where $E, F \in \M(A)$, 
should be implementable by a physical process $\phi : \V(A) \rightarrow \V(A)$, i.e., by an element of $\Proc(A)$. Moreover, this process 
should be reversible if the underlying process $\phi_E$ is reversible. 
So, suppose $x \in E \in \M(A)$, and let $\phi_{x} : \V(A_x) \rightarrow \V(A_x)$ be {\em any} 
(reversible) process (i.e., not only a symmetry) in $\Proc(A_x)$. We want this to be implementable by a (p-reversible) process $\phi : A \rightarrow A$ in $\Proc(A)$ such that 
$\phi(\beta) = \phi_{x}(\beta)$ for all $\beta \in F_x$, and 
$\phi(\delta_x) = \delta_x$.  %{\bf (This passage confused??)}%// $\phi^{\ast}(x) = x$.   
This suggests a %the following
}

Since we are considering physical processes $\phi \in \Proc(A)$ that need not be normalization-preserving (in particular, 
need not be associated with symmetries in $G(A)$), the following stronger version of the subspace property seems quite as well-motivated a the weaker version discussed above. %{\bf (Make a numbered definition?)}
\\

\noindent{\bf Definition B.6} Let $\Cat$ be a probabilistic theory in which all systems are sharp. Say that 
$\Cat$ has the {\em strong subspace property} (SSP) iff, for every system $A \in \Cat$ and every test $E \in \M(A)$, if $x \in E$, then (i) $A_x \in \Cat$ and 
(ii) any (reversible) process $\phi \in \Proc(A_{x})$ lifts to a (reversible) process in $\Proc(A)$ {\em fixing $\delta_{x}$}.\\ %{\blue: \bf [Do we also need point-separating?]}\\

{\blue Note that the significant difference here from the subspace property of Definition B.1 is the requirement 
that {\em all} processes on $A_x$, and not only symmetries, lift to processes on $A$, which can be taken 
to be reversible if the original processes are.}\\

\tempout{{\blue [REF 2:  `` the author should highlight more what the difference is from definition B.1"]}. 
%[OR: {\em fixing $E$}] [and reversible if $\phi$ is reversible?]
\noindent{\bf Definition B.1}  Say that a probabilistic theory $\Cat$ has the {\em subspace property} iff, for every 
$A \in \Cat$ and every $x \in X(A)$, 
\begin{mlist} 
\item[(i)]  $\M(A)_x$ separates points of $F_x$,  
\item[(ii)] The model $A_{x} := (\M(A)_{x}, F_{x})$ belongs to $\Cat$, and 
\item[(iii)] every symmetry in $G(A_x)$ extends to a symmetry $g \in G(A)$ with $gx = x$.\\
\end{mlist}
\\}

\tempout{\begin{quote}{\bf Strengthened Subspace Postulate (SSP):} Let $A$ be sharp. For every system $A \in \Cat$ and every test $E \in \M(A)$, if $x \in E$, then (i) $A_x \in \Cat$ and 
(ii) any (reversible) process $\phi : \V(A_x) \rightarrow \V(A_x)$ in $\Proc(A_{x})$ lifts to a (reversible) process 
$\V(A) \rightarrow \V(A)$ in $\Proc(A)$ {\em fixing $\delta_{x}$}. %[OR: {\em fixing $E$}] [and reversible if $\phi$ is reversible?]
\end{quote} 
}

\noindent{\bf Lemma B.7} {\em If $\Cat$ satisfies the SSP, every model in $\Cat$ has arbitrary reversible filters.}\\

\noindent{\em Proof:} A system of rank $1$ automatically has arbitrary filters: the only test has a single outcome, $x$, 
and there is just one state, $\delta_x$. The mapping $\delta_x \mapsto t_x \delta_x$ defines a p-reversible positive 
mapping on $\V(A) \simeq \R$.

Now suppose, for purposes of induction, that every system of rank $< n$ has arbitrary reversible filters. 
Let $E$ be a test, and let $\phi_{E} : \V(E \setminus \{x\}) \rightarrow \V(E \setminus \{x\})$ be a filter 
with $\phi(\delta_y) = t_y \delta_y$ for all $y \in E \setminus \{x\}$. Extend this to 
$\phi : \V(A) \rightarrow \V(A)$ fixing $\delta_x$. Now let $z \in E \setminus \{x\}$. Repeating the argument, 
find $\psi$ a filter on $A$ with $\psi(\delta_y) = \delta_y$ for all $y \in E \setminus \{x\}$ and {\blue $\psi(\delta_{x}) = t_x \delta_x$ with $t_x < 1$.}
(In other words, for $z \in E \setminus \{y\}$, we have $t_z = 1$ if $z \not  = x$ and $t_x$ the given value.) Composing, we have a positive mapping with 
\[(\psi \circ \phi)(\delta_z) = \psi(t_z \delta_z) = t_z \psi (\delta_z)  = t_z \delta_z\]
for $z \not = x$, and 
\[(\psi \circ \phi)(\delta_x) = \psi(\delta_x) = t_x \delta_x.\]
Thus, we have arbitrary p-reversible filters.  $\Box$ \\

In \cite{Wilce09}, it is shown that the existence of arbitrary reversible filters plus full symmetry implies the homogeneity 
of the state space. So we have \\

\noindent{\bf Proposition B.8} {\em Let $\Cat$ satisfy the SSP. If every $A \in \Cat$ is symmetric, then for every $A \in \Cat$, $\V(A)$ is homogeneous.}\\

\noindent{\em Proof:} By Lemma B.7 above, we have arbitrary reversible filters; by Lemma B.5 in previous subsection, every 
$A \in \Cat$ is fully symmetric. $\Box$ \\

The proofs of Theorem 1 and Corollary 1 imply that, in the presence of a spectral decompostion for states, sharpness, the existence of a weak conjugate, and the existence of arbitrary reversible filters are enough to secure the self-duality 
of $\E(A)$. Thus, Propositions B.3 and B.8 gives us a third route to Jordan models, to put alongside Corollaries 1 and 2 in the main body of this paper:\\

\noindent{\bf Corollary B.9} {\em Let $\Cat$ satisfy the SSP. If every model $A$ in $\Cat$ has a compact 
outcome-space $X(A)$, is symmetric and has a weak conjugate, 
then $A$ is homogeneous and self-dual, and hence, a Jordan model. }\\

\subsection{Jordan models and the SSP} 

The preceding results provide an alternative route from four operationally meaningful axioms --- sharpness, symmetry, the SSP and the existence of weak conjugates --- to euclidean Jordan algebras. The question 
remains whether all EJAs arise in this way.  In fact, the full symmetry enforced by Lemma B.5 significantly constrains the possibilities.\\

\noindent{\bf B.10 Lemma:} {\em Let $A = A(\J)$ be the Jordan model associated with a Jordan algebra $\J$. If $A$ is fully symmetric, then $\J$ is either simple, or a direct sum of one-dimensional Jordan algebras.} \\

\noindent{\em Proof:} Suppose $\J = \bigoplus_{i=1}^{n} \J_{i}$ where $\J_{1},...,\J_n$ are simple euclidean Jordan algebras, with $n \geq 2$.  Let $E_i$ be a Jordan frame for $\J_i$; then  
$E = \bigcup_{i=1}^{n} E_i$ is a Jordan frame for $\J$. Let $x_1 \in E_1$ and $x_2 \in E_2$. By full symmetry, there exists a symmetry $g \in G(A)$ such that 
$g x_1 = x_2$, $g x_2 = x_1$, and $gy = y$ for all $y \in E \setminus \{x_1, x_2\}$. Now let $p_i$ be the central projection associated with {\blue the $i$-th} summand, so 
that $p_1$ and $p_2$ are the central covers  of (the smallest central projections above) $x_1$ and $x_2$, respectively. Hence, $g(p_1)$ is the central cover of $gx_1 = x_2$, i.e, $g(p_1) = p_2$, and 
similarly $g p_2 = p_1$. 
But now if $y \in E_1$ is any point other than $x_1$, we have $y = gy \leq gp_1 = p_2$, which are impossible, as $y$ is orthogonal to $p_2$.  So $E_1 = \{x_1\}$, and  $\J_1$ is one-dimensional.  Since $\J_1$ is arbitrary, all summands are 
one-dimensional. $\Box$ \\

Thus,  a probabilistic theory satisfying the hypotheses of Corollary B.9 will comprise only simple Jordan models and classical systems. In particular, such a theory cannot accommodate superselection rules. 
This is also true, however, of earlier reconstructions based on versions of the subspace axiom that assume all systems of the same information capacity to be isomorphic.  (If the theory is {\em monoidal}, roughly meaning that 
it allows for the formation of composite systems --- and meaning more precisely that the category in question 
has a symmetric monoidal structure, satisfying some reasonable constraints on how this interacts with the theory's probabilisic apparatus; see \cite{BW12} for details --- then there are further constraints. These matters are discussed in \cite{BGW}.  For a discussion of how monoidal ``process theories" in the sense of \cite{Abramsky-Coecke, Coecke-Kissinger} can be represented as probabilistic theories in the sense of this paper, see \cite{Shortcut}.)
 
%{\blue \bf [REF 2 wants more commentary on the connection w/ monoidal categories in the main text. (Could reference shortcut?)]}} 

We now show that the assumptions discussed in this Appendix imply no further constraints on Jordan models. 
To this end, it will be enough to exhibit a probabilistic theory $\Cat$ containing all simple Jordan models, 
and assigning to each such model $A$ a monoid $\Proc(A)$ of allowed processes in such a way that $A$ is 
symmetric and $\Cat$ enjoys the SSP.  \\

\noindent{\bf Proposition B.11:} {\em Let $\Cat$ be the probabilistic theory consisting of all full, simple euclidean Jordan models, with processes on a given model $A(\J)$ consisting of composites of maps of the form $U_{a}$, $a \in \J$
\footnote{The {\em structure group}, $\Gamma(\J)$, of a euclidean Jordan algebra $\J$ consists 
of all non-singular positive mappings $\phi : \J \rightarrow \J$ such that $U_{\phi}(a) = \phi U_{a} \phi^{\ast}$ for 
every $a \in \J$. A theorem of Koecher \cite{Koecher} shows that the connected component of the identity 
in $\Gamma(\J)$ consists exactly of composites of mappings of the form $U_{a}$. In other words, in this theory, $G(A)$ is precisely the connected identity component of $\Gamma(\J)$.}. Then every model in $\Cat$ is symmetric, and satisfies the SSP.} \\

\noindent{\em Proof:} 
Let $A = A(\J) = (\M(\J),\Omega(\J))$ be the Jordan model associated with an EJA $\J$. Thus, 
$\M(\J)$ is the set of Jordan frames, and $X(\J)$ the set of primitive idempotents, of $\J$. 
If $\J$ is simple, symmetry (indeed, full symmetry) of $A$ under $G(A)$ follows from (\cite{FK}, IV.2.5)\footnote{Or,  
more exactly, from the proof thereof. For any two Jordan-orthogonal primitive idempotents $x,y$ there exists 
a symmetry --- an element $s \in A$ with $s^2 = u$ --- such that $U_{s}(x) = y$ (\cite{FK}, IV.2.4).  This, plus 
an induction, allows one to construct an automorphism of the form $U_{a}$ taking one Jordan frame to another.} 
It follows easily that $G(A)$ acts transitively on Jordan frames even if $\J$ is not simple (since each such frame 
is the disjoint union of frames {\blue chosen} from each simple summand). 
Let $p \in P(\J)$. Then (\cite{Alfsen-Shultz}, Prop. 1.38, Lemma 1.39, Prop. 1.43, Prop.  2.32 and remarks preceding latter), $\J_p := U_{p}(\J)$ is a hereditary Jordan subalgebra, 
meaning that if $0 \leq a \leq b \in \J_p$, then $a \in \J_p$ as well. The unit of $\J_p$ is $p$.  It is important to note here that $(\J_p)_+ =  \J_+ \cap \J_p$ (that is, an element of $\J_p$ is positive in $\J_p$ iff it is positive in $\J$). This follows from spectral theory; see \cite{Alfsen-Shultz} Prop. 1.22 and 
subsequent discussion. It follows that if $a, b \in \J_p$, $a \leq b$ in $\J_p$ iff $a \leq b$ as elements of $\J$.
One can also show (\cite{Alfsen-Shultz}, Lemma 1.45) that if $a \in \J_p$ and $b \in \J_{p'}$, then 
$a \dot b = 0$. 

Now let $e$ be a primitive idempotent in $\J$ that happens to lie in $\J_p$: then $e$ is an idempotent in $\J_p$ as well. 
If $f$ is another nonzero idempotent in $\J_p$ with $f \leq e$, then $f$ is still idempotent in $\J$ and, by remarks 
above, $f \leq e$ in $\J$, whence, as $e$ is minimal, $e = f$. Thus, $e$ is primitive in $\J_p$. 
Conversely, let $e$ be primitive idempotent in $\J_p$. Then $e$ is still idempotent in $\J$. If 
$f$ is an idempotent of $\J$ with $0 < f \leq e$ in $\J$, then $f \in \J_p$, since the latter is hereditary. 
As $e$ is primitive in $\J_p$, 
$f = e$. Thus, $e$ is primitive in $\J$. 

This shows that $X(\J_p) = \J_{p} \cap X(\J)$.  It follows that $\M(\J_p)$ is the set of all 		
sets of primitive idempotents of $\J$ summing to $p$. Now let $p = u - x$ where $x$ is a primitive idempotent of $\J$. Then $F \in \M(\J_p)$ iff $F \cup \{x\} \in \M(\J)$. In other words, $\M(\J_p) = \M(A)_{x}$ in the notation of 
Definition B.1. 

It remains to show that any (p-reversible) process in $\Proc(A_{x})$ extends to a (reversible) process in 
$\Proc(A)$ leaving $x$ fixed. It suffices to show this holds for processes of the form $U_a$, with $a \in \J_p$. 
Since $x \dot b = b \dot x = 0$ for all $b \in \J_p$, a straightforward 
calculation then shows that $U_{a + x}b = U_{a}b$ for all $b \in \J_p$, and that $U_{a + x} x = x$. Hence, 
$U_{a + x}$ is the desired extension of $U_{a}$. Note that if $U_{a}$ is invertible, then $a$ is invertible 
in $J_{p}$. By Lemma A.4 (b), there exists some $b \in \J_p$ with $a \dot b = p$ and $a^2 \dot b = a$. 
Since $x \dot b = 0$ for all $b \in \J_{p}$, we have 
\[(a + x) \dot (b + x) = a \dot b + x^2 = a \dot b + x = p + x = u,\] 
and 
\[(a + x)^2 \dot (b + x) = (a^2 + x) \dot (b + x) = a^2 \dot b  + x^2   = a + x.\] 
Invoking Lemma A.4 (b) again, this implies 
that $b + x$ is the inverse of $a + x$ in $\J$, whence, by Lemma A.4 (c), $U_{a + x}$ is invertible. $\Box$
% $a \in \J_{+}^{\int}$, 
%... NEED: $a + x$ invertible as well!!}$\Box$ 

\end{document}